\documentclass{article}

\usepackage{arxiv}
\usepackage{comment} 
\usepackage[utf8]{inputenc} 
\usepackage[T1]{fontenc}    
\usepackage{hyperref}       
\usepackage{url}            
\usepackage{booktabs}       
\usepackage{amsfonts}       
\usepackage{amsmath}        
\usepackage{amsthm}         
\usepackage{nicefrac}       
\usepackage{microtype}      
\usepackage{lipsum}		
\usepackage{graphicx}
\usepackage{natbib}
\usepackage{doi}
\usepackage{amsmath}
\usepackage{float}
\usepackage{stmaryrd}

\usepackage[table]{xcolor}
\definecolor{lightblue}{rgb}{0.8,0.92,1}  
\definecolor{darkblue}{rgb}{0.0,0.0,0.5}  

\usepackage[shortlabels]{enumitem}

\usepackage{multirow}
\usepackage{breqn} 
\usepackage{mathrsfs}

\title{Towards Explainable Automated Data Quality Enhancement without Domain Knowledge}

\author{Djibril~Sarr \\
	LAGA, UMR  7539 CNRS\\
	Université Sorbonne Paris Nord\\
	99 Av. Jean Baptiste Clément, 93430 Villetaneuse\\
	FBH Associés \\
	11 Rue du 4 septembre, 75002 Paris\\
	\texttt{sarr@math.univ-paris13.fr} \\
	\texttt{djibril.sarr@fbh-associes.com} \\
}

\renewcommand{\headeright}{}
\renewcommand{\undertitle}{}
\renewcommand{\shorttitle}{Towards Explainable Automated Data Quality Enhancement without Domain Knowledge}

\hypersetup{
pdftitle={Towards Explainable Automated Data Quality Enhancement without Domain Knowledge},
pdfsubject={Data quality},
pdfauthor={Djibril SARR},
pdfkeywords={Data quality, statistical outliers, typographical errors, logical errors},
}

\begin{document}
\maketitle

\begin{abstract}
In the era of big data, ensuring the quality of datasets has become increasingly crucial across various domains. We propose a comprehensive framework designed to automatically assess and rectify data quality issues in any given dataset, regardless of its specific content, focusing on both textual and numerical data. Our primary objective is to address three fundamental types of defects: absence, redundancy, and incoherence. At the heart of our approach lies a rigorous demand for both explainability and interpretability, ensuring that the rationale behind the identification and correction of data anomalies is transparent and understandable. To achieve this, we adopt a hybrid approach that integrates statistical methods with machine learning algorithms. Indeed, by leveraging statistical techniques alongside machine learning, we strike a balance between accuracy and explainability, enabling users to trust and comprehend the assessment process. Acknowledging the challenges associated with automating the data quality assessment process, particularly in terms of time efficiency and accuracy, we adopt a pragmatic strategy, employing resource-intensive algorithms only when necessary, while favoring simpler, more efficient solutions whenever possible. Through a practical analysis conducted on a publicly provided dataset, we illustrate the challenges that arise when trying to enhance data quality while keeping explainability. We demonstrate the effectiveness of our approach in detecting and rectifying missing values, duplicates and typographical errors as well as the challenges remaining to be addressed to achieve similar accuracy on statistical outliers and logic errors under the constraints set in our work.
\end{abstract}


\section{Introduction}
\label{sec:Intro}
Data assumes an escalating significance in contemporary business operations. Across diverse sectors such as commerce, entertainment, medicine, and the oil industry, data emerges as a potent catalyst for augmenting business efficacy \citep{DATA_1}. An abundant reservoir of qualitative data, coupled with Artificial Intelligence (AI), statistical methodologies, and probabilistic approaches, facilitates the discernment of intricate behavioral patterns or associations that would be arduous for humans to discern. However, the efficacy of these methodologies depends on the availability of high-quality data, often necessitating voluminous datasets. Notably, the preparatory phase, which entails rendering data compatible with analytical algorithms, has long been estimated to consume an important part of the process of exploiting the data. Two decades prior, \cite{DATA_2} already estimated the proportion of time spent to be over 50\% of the overall processing time. Presently, notwithstanding the advancements in analytical techniques, the increasing volume and intricacy of data sources have exacerbated the challenges associated with data preparation \citep{garcia2016big}. This critical aspect, known as data preprocessing, encompasses a suite of techniques executed prior to the extraction of actionable insights from the data, constituting a pivotal facet of data exploitation. Due to the inherent anomalies in data, initiating data mining endeavors directly is impractical without addressing inconsistencies and redundancies. Furthermore, the exponential surge in data generation rates across multiple sectors necessitates the adoption of increasingly sophisticated analytical mechanisms \citep{han2022data, zaki2014data}. In light of constraints pertaining to time and human resources, there exists a pressing imperative to automate and streamline this preparatory phase. Central to this process is the identification and rectification of erroneous data values. They constitute a crucial component of the data cleansing process (refer to, for instance, Section 3.2 of \cite{datacleansing}).

For the reasons outlined above, it is imperative for practitioners to possess efficient tools for measuring and enhancing data quality. This study focuses on optimal strategies for providing automated tools for this phase. We also assert that without explainability and interpretability, the effectiveness of data quality assessment algorithms is compromised, as users may lack confidence in the reliability of the results. For instance, if an entry is deemed invalid due to the detection of a statistical outlier on a specific line of a data table, but the problematic field(s) cannot be precisely identified by the algorithm, it becomes challenging to fully rely on the algorithm for making informed decisions. We propose a framework comprising five major steps that can automatically identify defects in a dataset without prior knowledge of its contents. This is accomplished while ensuring the ability to explain and justify each conclusion reached by our framework. This feature enables the framework, which consists of a series of algorithms, to provide native corrections for all errors identified in a given dataset. Returning to the earlier example, the ideal framework should not only identify the presence of a statistical outlier but also pinpoint the problematic field and propose a correction. Our framework is applicable to any table containing numerical values, text, or both. It addresses three types of defects that can affect data: absence, redundancy, and inconsistency. The first two steps of the five-steps algorithm we introduce deal with the first two types of errors and the last three steps all handle inconsistencies as we consider statistical outliers, typographical errors and logic errors. While the first two defects may be self-explanatory, indicating respectively missing and duplicated values, the third is less straightforward. In our framework, inconsistencies can impact any type of observation, whether numerical or textual, occurring when an input in a given field appears abnormal relative to others in the same field. While Artificial Intelligence (AI)-based techniques may offer potentially superior performance, they often lack the transparency necessary for thorough understanding and validation of results. Therefore, the algorithms within our comprehensive framework attempt to combine AI-based techniques and statistical results to identify and correct defects in given datasets. This characteristic, specifically its outcome of being able to explain all results, is the main distinction between our work and existing literature.

The literature contains several algorithms for analyzing the validity of a data set. As in this work, many of them leverage Machine-Learning (ML), Deep-Learning (DL) and statistical methods. 
Considering specifically the handling of missing values, they do not constitute per se a tough challenge as they can be identified in pretty straightforward ways. However, in the process of handling them, two challenges arise. The first one is analyzing the rightfulness of the missing data. Indeed, a missing value does not necessarily designate a defect in the dataset as it can, in various situations, be a legitimate and information-providing entry. As developed in \cite{sainani2015dealing}, the most straightforward and commonly employed approach to managing missing data involves excluding incomplete observations entirely from the analysis. That solution, while very easily applicable, is drastic. As already stated, it might delete the valuable information provided by the missing value, but it obviously also discriminates all the other valid entries. The author also presents the very common solution, which is to impute the missing value with a representative value of the field. It can be the average value (for instance, if it is heights of people), it can be a \textit{local} average, for instance the weight of the males in a specific study, and so on and so forth. Other researchers have instead used ML and DL to handle missing values. For instance, \cite{savarimuthu2021unsupervised} imputation method on time series uses an iterative imputation algorithm that clusters univariate time series data, taking into account the data's trend, seasonality, and cyclical patterns. Then within the cluster, the authors employ a similarity-based nearest neighbor imputation method within each cluster to fill in missing values. More recently, \cite{kachuee2020generative} used a generator network to generate imputations that a discriminator network is tasked to distinguish; this method allows also to properly estimate the distribution of the targets. While methods using AI are shown by their authors to be very efficient, they hardly provide the level of explainability that we aim for. For these reasons, in section \ref{sec:Abs}, we will only use a classic imputation method for handling missing numerical values, ensuring explainability and interpretability. However, for character entries, when addressing logic errors in Subsection \ref{sec:LogicErr}, we will employ ML algorithms for imputation while maintaining transparency. Our strategy will guarantee that legitimate missing values will not be discriminated, thanks to the subsequent application of the logic error detection algorithm. Our approach also guarantees that every choice will be explainable and interpretable.

Research has also explored the combination of neural networks and statistics specifically for outlier detection. For example, Small et al. \cite{smallaibased2018} trained a neural network to replicate a specific field within their dataset to identify outliers. They compared the neural network's output with the actual values and employed statistical indicators similar to those discussed in Subection \ref{sec:outliers}. The authors calculated the difference between the output and actual values of the field under consideration, constructing a tolerance interval based on the distribution of errors between the two values. If the error fell outside the tolerance interval, the actual value was deemed to have been incorrectly attributed. Therefore, their approach necessitates building and training a neural network for identifying statistical outliers, which can be costly, particularly for high-dimensional datasets. The reliability of this method relies on the neural network's performance, as the entire approach depends on its ability to accurately reproduce actual values. Another commonly explored approach in the literature is the use of DBSCAN, Density-Based Spatial Clustering of Applications with Noise \citep{dbscan1,dbscan2}. However, the challenge with this method lies in selecting the parameters $\epsilon$, which defines the maximum distance between points in the same neighborhood, and $minPts$, the minimum number of points needed to form a dense region. This parameter selection issue has been extensively discussed in the literature \citep{limitdbscan1, limitdbscan2}. Moreover, considering our goal of achieving maximum interpretability and explainability, it is essential to note that DBSCAN is designed for clustering and density-based outlier detection in multi-dimensional datasets, making it challenging to maintain explainability while identifying outliers, particularly as defined in Section \ref{sec:Definitions}. 

Regarding typographical errors, they have long been a focus of researchers, particularly in text analysis. Early scholars, such as those referenced in \cite{typos1975}, were among the first to address this issue. As discussed in Subection \ref{sec:Typos}, their approach, similar to ours, drew insights directly from the dataset, enabling the algorithm to adapt dynamically to the text under scrutiny. For example, they utilized statistical analysis of word trigrams within their documents. In contrast, our method, as explained later, utilizes word frequencies extracted from the table we aim to rectify. Our typographical error detection and correction algorithm incorporates machine learning techniques. This integration of AI and text analysis is well-documented; previous researchers have employed machine learning and deep learning methods to rectify typographical errors, often leveraging contextual cues or domain-specific knowledge (e.g., \cite{huang2013automotive}). In our study, we opt for unsupervised machine learning approaches, particularly clustering algorithms, to avoid dependence on domain-specific knowledge and to expedite the identification of typographical errors. Furthermore, much of the existing literature in this field relies on external datasets. For instance, one step in the algorithm proposed by \cite{googletypo} involves cross-referencing each word with Google Web 1T 5-Gram extensive words' unigrams dataset. The algorithm we introduce doesn't require external information. It can however be used when desired to increase the speed of the process. The application we will provide will solely use information that is provided within the data sets. 

Analyzing and correcting logic errors is not as frequently addressed in the literature as the other types of errors we have previously discussed, particularly with the definition provided in Subsection \ref{sec:outliers}. However, as we will see in details, identifying logic errors can be likened to recognizing anomalous behavior across an entire observation (taking into account all fields of the data table) in comparison to others. This can include tasks such as detecting credit card fraud or identifying cancer cells. One of the model families commonly employed for this purpose is rule-based algorithms, as demonstrated by \cite{diabetes} in diagnosing Type II diabetes. Similarly, this is the model family we will employ for logic error detection. Nonetheless, numerous alternative solutions also exist. For instance, in \cite{variationalautoencoders}, variational auto-encoders are utilized to detect seasonal anomalies in key metrics (such as the number of views, online users, and orders) of web applications. Alternatively, Support Vector Machines (SVMs) have been explored, as demonstrated in \cite{creditcard} for credit fraud detection. A common characteristic among all these approaches, including ours, is their reliance on utilizing all available fields for a specific observation. This requirement was not necessarily present in the handling of other defects discussed in this document. 

After conducting an extensive review of the current state of the art, it appears that this paper aims to address a notable gap in the data cleansing literature: the need for an explainable, interpretable, and automatic solution. While numerous efficient solutions are available, they do not have the dual constraints emphasized in our work. By applying our approach to a publicly available agricultural equipment sales data set, we illustrate its effectiveness, in detecting and rectifying missing values, duplicates and typographical errors while being able to keep a maximal explainability and without using domain knowledge. Similarly, we discuss the challenges remaining to be addressed to achieve similar accuracy on statistical outliers and logic errors under the same constraints.

In Section \ref{sec:Definitions} of this document, we present fundamental definitions that underpin the selection of algorithms and strategies. These definitions primarily address the concepts of data validity and the explainability of an algorithm, and consequently, its results. Following this, in Section \ref{sec:Data}, we introduce the dataset on which we will apply the framework. This will allow us to provide concrete illustrations when presenting the algorithms. Then, Section \ref{sec:Framework} will be the core of this work as we introduce the framework. Initially, we offer a broad overview of the process, followed by a detailed exposition of each of its five steps. Subsequently, in Section \ref{sec:Results}, we present the results obtained after applying the framework to the dataset.

\section{Definitions}
\label{sec:Definitions}
We begin by presenting key definitions that will guide the selection of algorithms and shape the focus of the framework that we will introduce. Specifically, we will define what we will consider in this work to be valid data and we will provide a contextualised definition of explainability and interpretability. It is worth emphasizing that these definitions might not achieve unanimous agreement in the literature, as they do not always capture clear and universally objective concepts. 

In what follows, a \textit{data set} \( D \) is defined as a finite collection of $n$ records \( \{ r_1, r_2, \ldots, r_n \} \), where each record \( r_i \) is a tuple consisting of \( m \) attributes. Formally, we can represent the data set as:
\[
D = \{ r_1, r_2, \ldots, r_n \},
\]
where \( r_i = (a_{i1}, a_{i2}, \ldots, a_{im}) \) for \( i = 1, 2, \ldots, n \). Each attribute \( a_{ij} \) is an element of a predefined attribute domain \( A_j \). This formalization allows us to consider a data set as a structured collection of multi-dimensional data points, where the dimensionality is determined by the number of attributes \( m \). Note that, any element $a_{ij}$ can be a numerical value as well as a text value.

\subsection{Valid data}
In the literature, data quality is frequently characterized by the presence or absence of defects. According to this perspective, data is considered to be of high quality if it is devoid of such defects or if their impact is minimal. This conceptualization can be seen for instance in the work of \cite{schelter2018automating} and \cite{CITE_12}. We will also adhere to this  approach in this paper. It is also important to note that throughout this document, \textit{the data} is not defined explicitly when unambiguous. It encompasses both individual observations which corresponds to a specific column of a row within a data array and objects of dimensions greater than one denoted as $p > 1$. They refer to a row or subsets thereof within a data array. Also, obviously the observations can be text as well as numerical values. Let us now define the defaults that will form the basis of the framework. 

\begin{enumerate}
    \item \textbf{Absence: }\\
    This refers to the absence of data or its constituent elements, which can arise during the data collection process due to technical failure, error, or insufficient information \citep{CITE_14}. In such instances, the data is deemed invalid.
    
    \item \textbf{Redundancy: }\\
    Data (where $p \geq 1$) that is replicated across multiple entries is regarded as invalid and is often denoted as duplicates in the literature.
    
    \item \textbf{Inconsistency: }\\
    We define inconsistency as the lack of agreement between a data item and other observed data. Consequently, it encompasses several types of defects:
    
    \begin{enumerate}
      \item \textbf{Statistical outliers:}\
        These are data points that are significantly and unjustifiably distant from other observations. An abnormal value may result from inconsistency or aberrant behavior during the measurement process \citep{CITE_17, CITE_18}.
        
         \item \textbf{Typographical errors:}\
        These errors encompass all text-related issues, including typing errors (e.g., "France" vs. "Fran\underline{g}e") and formatting discrepancies (e.g., "Citroën" vs. "Citro\underline{Ã?}n").
                
        \item \textbf{Logical errors:}\
        These errors, more challenging to define, denote incompatibilities between different variables (columns) within a dataset.
    \end{enumerate}
\end{enumerate}

Data will, therefore, be considered valid only when it exhibits none of these defects. The forthcoming framework will be  crafted to detect and address these varied defaults. It will strive to rectify such anomalies where feasible, all the while prioritizing the preservation of explainability and interpretability of both the algorithms used and their results.

\subsection{Explainability and Interpretability}
The lack of explainability and Interpretability and the \emph{black box} character it entails is one of the main arguments against a more widespread and systematic use of ML or DL techniques \citep{CITE_20}. Therefore, one of our focal point is to uphold these principles of explainability and interpretability to the fullest extent we can realize within this framework.

To facilitate our comprehension of the concept of interpretability, let us begin with a more literary perspective. According to the \emph{Larousse}, the verb \textbf{interpret} refers to the act of "seeking to make a text, an author intelligible, explaining them, commenting on them.". In the context of statistical algorithms and machine learning, as emphasized by \cite{CITE_21}, this intelligibility is directed towards humans. An algorithm is considered interpretable if the results it generates can be articulated in terms understandable to humans. Meanwhile, the concept of \textbf{explainability} is more commonly encountered and appears inherently more intuitive. However, the literature does not always converge on a formal definition of this concept within an algorithmic context. Beginning again with a literary perspective, \emph{Larousse} defines explainability as "making someone understand (a question, an enigma), clarifying them by providing the necessary elements." \citep{CITE_22} emphasizes that every stage of the clarification process contributes to explainability. To provide a comprehensive view, \citep{CITE_23} defines explainability (particularly in a deep learning context) as the combination of interpretability and transparency. Transparency, in this context, refers to the degree to which an explanation renders a specific result acceptable. It is important to note that acceptance of the result does not necessarily imply that the practitioner perceives it as right or wrong, but rather, that the path leading to that specific result is clear.

Drawing from both the literary perspective and insights from the literature, it becomes evident that the concepts of interpretability and explainability play pivotal roles in making sense of algorithmic outputs, especially within the context of statistical algorithms and ML. We can now, in our context provide the following definitions.

\begin{enumerate}
    \item \textbf{A result that can be explained and interpreted will refer to: }
    \begin{enumerate}
        \item A result for which we can understand the path, regardless of its correctness. That is, we must be capable of elucidating the inputs, outputs, rules, etc., through which a defect (as defined in the previous section) has been identified.
        \item A result whose boundaries can be recognized and comprehended is imperative. That is, one must be able to understand, explain, and deliberate on the scenarios where a result, whose path has already been understood, is meaningful or not.
        
        In addition to these two elements that respectively address explainability and interpretability, we introduce a third equally significant constraint.
        
        \item A result that we can clearly identify and locate. That is, within a data quality context, if a particular row is flagged as problematic, we must be able to identify which columns are deemed incorrect.
    \end{enumerate}
    
    \item \textbf{An algorithm that can be explained and interpreted is:}
    \begin{enumerate}
        \item An algorithm whose results can be explained and interpreted.
    \end{enumerate}
    
\end{enumerate}

For the sake of conciseness, we will refer to the duality of explainability and interpretability simply as explainability.

\section{Data set for the application of the framework}  
\label{sec:Data}

In the following sections, we will introduce each of the algorithms comprising the framework outlined in Section \ref{sec:Framework}. To provide clarity on how these algorithms operate, we propose starting by introducing the dataset. This approach will enable us to reference the dataset when necessary to illustrate an algorithm's function.

The data set used in this work is build upon the Blue Book for Bulldozers data set publicly available\footnote{\url{https://www.kaggle.com/c/bluebook-for-bulldozers/data}}. The data set is made available by French agency AMIES (Agency for Mathematics in Interaction with Industry and Society\footnote{Agence pour les Mathématiques en Interaction avec l'Entreprise et la Société}) for their 2021 competition\footnote{The work presented in this article won the first prize of the data quality 2021 competition (\href{https://challenge-maths.sciencesconf.org/resource/page/id/1}{https://challenge-maths.sciencesconf.org/resource/page/id/1}).}. 

The database contains 100,000 observations described by 53 parameters briefly defined in the table \ref{tab:data}. This table categorizes various attributes of auction machinery, grouping them primarily under unique identifiers, descriptive information, configuration details, and location-specific data. Key identifiers such as SalesID, MachineID, ModelID, datasource, and auctioneerID are crucial for tracking and analyzing sales transactions. Descriptive attributes like YearMade, MachineHoursCurrentMeter, and UsageBand provide insights into the machine's age, usage level, and operational status. The table also extensively details machine configurations, highlighting features such as Drive\_System, Enclosure, Forks, and numerous others that specify the machine's physical and operational setup, which is critical for potential buyers to assess the machinery’s capabilities and condition. Furthermore, geographical and market segmentation is addressed through variables like State and ProductGroupDesc, which help in understanding the categorization of the machinery. This structured data arrangement aids in the comprehensive analysis and valuation of the machinery at the point of sale. Table \ref{tab:errs} below summarizes all the types of errors that we will be identifying and correction in this paper. Obviously, none of the information provided by Tables \ref{tab:errs} or \ref{tab:data} will be used in the algorithms.

\begin{table}[H]
    \centering
    \begin{tabular}{|p{4.25cm}|p{11.5cm}|}
        \hline
        \textbf{Variables } & \textbf{Description} \\ 
        \hline
        SalesID & Unique identifier of a particular auction machine sale.\\
        \hline
        MachineID & Identifier of a particular machine. \\
        \hline
        ModelID & Model identifier. \\
        \hline
        datasource & Unique machine identifier.\\
        \hline
        auctioneerID & Identifier of a particular auctioneer.\\
        \hline
        YearMade & Machine manufacturing year.\\
        \hline
        MachineHoursCurrentMeter & Current machine usage in hours at time of sale. \\
        \hline
        UsageBand & Value (low, medium, high) calculated by comparing the hours of this particular sales machine with the average usage of the base model.\\
        \hline
        Saledate &	Sale date.\\
        \hline
        Saleprice &	Cost in USD.\\
        \hline
        fiModelDesc &	Description of a unique model of machine.\\
        \hline
        fiSecondaryDesc	& fiModelDesc disaggregation\\
        \hline
        fiModelSeries &	fiModelDesc disaggregation\\
        \hline
        fiModelDescriptor &	Disaggregation of fiModelDesc \\
        \hline
        ProductSize &	Size class grouping for a product family.\\
        \hline 
        ProductClassDesc &	Description of the 2nd level hierarchical grouping.\\
        \hline
        State &	U.S. state in which the sale took place.\\
        \hline
        ProductGroup & Identifier for first-level hierarchical grouping.\\
        \hline
        ProductGroupDesc &	Description of high-level hierarchical grouping.\\

        \hline
        Drive\_System & Machine setup 1.\\
        \hline
        Enclosure & Machine configuration - whether the machine has an enclosed cab or not ?\\

        \hline
        Forks & Machine configuration - lifting accessory.\\
        
        \hline
        Pad\_Type &	Machine configuration - type of tread on a tracked machine.\\
        \hline
        Ride\_Control & Machine configuration - optional feature on loaders.\\
        
        \hline
        Stick &	Machine configuration - control type.\\
               \hline
        
        Transmission &	Machine configuration - describes the transmission type.\\
        \hline
        Turbocharged &	Machine configuration - naturally aspirated or turbocharged engine.\\
        \hline
        Blade\_Extension & Machine configuration - standard blade extension.\\
        \hline
        Blade\_Width &	Machine configuration - blade width\\
        \hline
        Enclosure\_Type &	Machine configuration - does the machine have an enclosed cab or not?\\
        \hline
        Engine\_Horsepower &	Machine configuration - motor power.\\
        \hline
        Hydraulics & Machine configuration - type of hydraulics.\\
        \hline
        Pushblock &	Machine configuration - option.\\
        \hline
        Ripper &	Machine configuration - tool attached to the machine to work the ground.\\
        \hline
        Scarifier &	Machine configuration - tool attached to the machine for soil conditioning\\
        \hline
        Tip\_control &	Machine configuration - type of blade control.\\
        \hline
        Tire\_Size &	Machine configuration - primary tire size.\\
        \hline
        Coupler & Machine configuration - type of machine interface.\\
        \hline
        Coupler\_System	& Machine configuration - machine interface type.\\

        \hline
        Grouser\_Tracks	& Machine configuration - describes ground contact interface.\\
        \hline
        Hydraulics\_Flow &	Machine configuration - describes the ground contact interface.\\
        \hline
        Track\_Type &	Machine configuration - type of tread on a tracked machine.\\
        \hline
        Undercarriage\_Pad\_Width	& Machine configuration - width of track treads.\\
        \hline 
        Stick\_Length	& Machine configuration - length of machine digging tool.\\

        \hline
        Thumb & Machine configuration - accessory used for input.\\
        \hline
        Pattern\_Changer &	Machine configuration - operator control configuration can be adapted to the user.\\
        \hline
        Grouser\_Type &	Machine configuration - type of tread on a tracked machine.\\
        \hline

        Backhoe\_Mounting &	Machine configuration - optional interface used to add an excavator accessory.\\
        \hline
        Blade\_Type	& Machine configuration - describes the blade type. \\
        \hline
        Travel\_Controls &	Machine configuration - describes the operator control configuration.\\
        \hline
        Differential\_Type & Machine configuration - differential type.\\
        \hline
        Steering\_Controls	& Machine configuration - describes the operator control configuration.\\
        \hline
    \end{tabular}
    \label{tab:data}
    \vspace{0.3cm}
    \caption{Fields in the data set}
\end{table}
\vfill

The data set is altered with different types of errors.

\begin{table}[h]
\centering
\begin{tabular}{|c|c|c|c|}
\hline
\textbf{Correspondence} & \textbf{Types of error} & \textbf{Number of cases} & \textbf{Subtotal} \\
\hline
Redundancy & Duplicates & 20 & \multirow{1}{*}{20} \\
\hline
Absence & Missing Value & 161 & \multirow{1}{*}{161} \\
\hline
Inconsistency (Outliers) & Aberrant Value & 200 & \multirow{1}{*}{200} \\
\hline
\multirow{3}{*}{Inconsistency (Typography)} & Entry Error & 100 & \multirow{3}{*}{200} \\
\cline{2-3}
 & Uppercase & 50 & \\
\cline{2-3}
 & Lowercase & 50 & \\
\hline
\multirow{5}{*}{Inconsistency (Logic)} & Wrong Category & 25 & \multirow{5}{*}{450} \\
\cline{2-3}
 & Incoherent Machine & 100 & \\
\cline{2-3}
 & Incoherent Drive System & 200 & \\
\cline{2-3}
 & Incoherent Product Group Description & 100 & \\
\cline{2-3}
 & YearMade $>$ saledate & 25 & \\
\hline
\multicolumn{3}{|c|}{\textbf{Total}} & 1031 \\
\hline
\end{tabular}
\vspace{0.3cm}
\caption{Summary of error types and correspondence}
\label{tab:errs}
\end{table}

Now that the dataset is properly introduced, we can proceed with presenting the algorithms within the data quality measurement and enhancement framework.

\section{Automatic data quality enhancement framework}  
\label{sec:Framework}

\begin{figure}[H]
    \centering
    \includegraphics[ width = 16.5cm, angle=0]{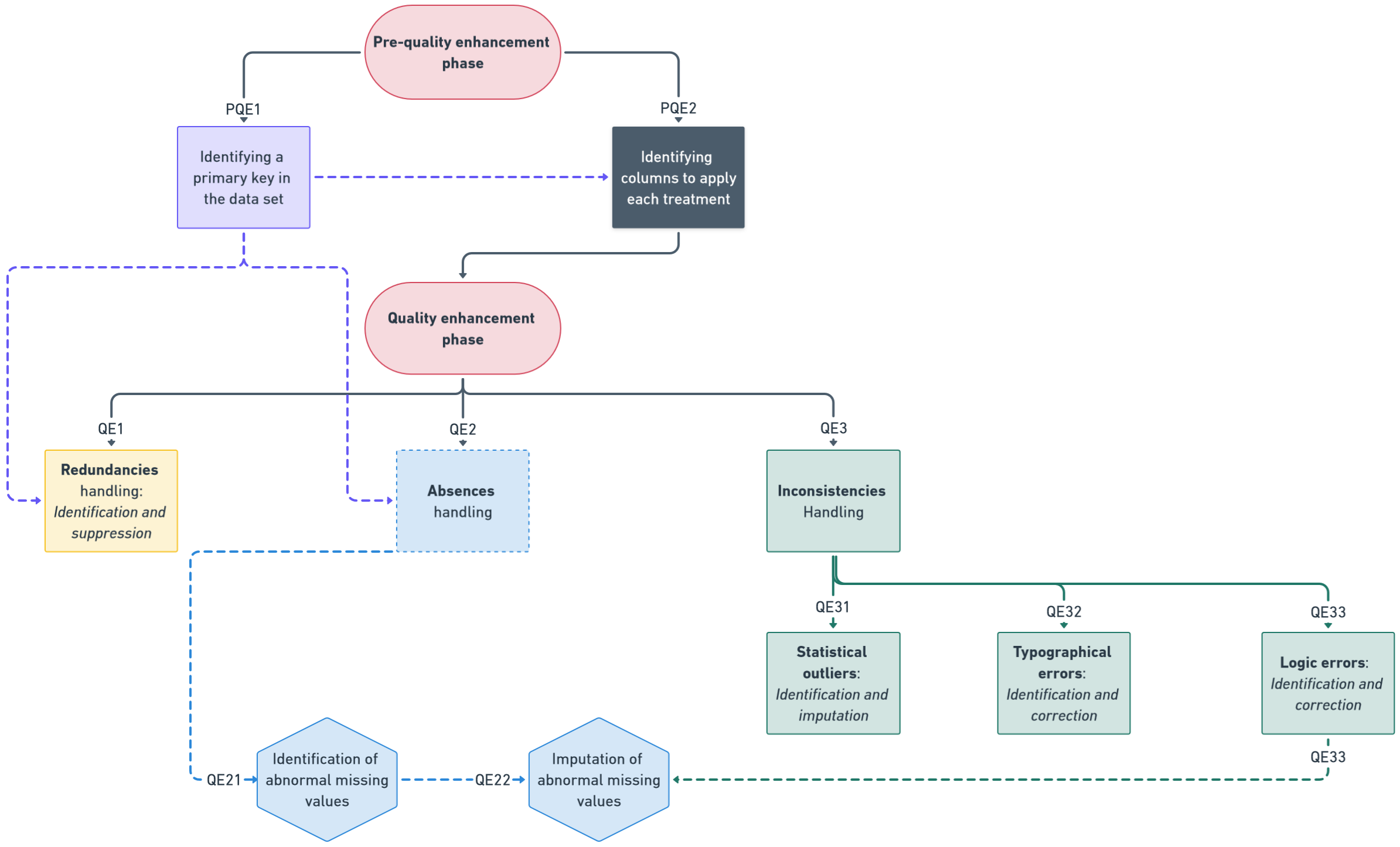}
    \caption{Invalid data detection framework}
    \label{fig:framework}
\end{figure}

The framework illustrated in Figure \ref{fig:framework} consists of two phases: one preceding the commencement of actual quality enhancement (Pre-Quality Enhancement, PQE phase), and the other constituting the actual data quality enhancement process (Quality Enhancement Phase, QE phase). Initially, the PQEP involves identifying a primary key in the dataset (PQE1) and determining the columns that will receive each of the required treatments (PQE2). These two steps are fundamental in enabling an automated process that does not rely on domain knowledge. Following this preparation, the QEP targets the three key areas that were mentioned earlier: 

\begin{itemize}
    \item \textbf{Redundancies handling} encompasses the identification and elimination of duplicate entries (QE1). As depicted in the flowchart, this step necessitates the utilization of the primary key identified in the PQE phase.
    \item \textbf{Absences Handling} concentrates on the imputation of abnormal missing values to preserve data integrity (QE12). This step is divided into two actions: firstly, the identification of abnormal missing values, and secondly, their imputation. Indeed, as explained in Subsection \ref{sec:Abs}, the process of identifying missing values and their imputation is not simultaneous. Similar to the preceding step, handling absences will also leverage the key identified during step A1 of the PQE phase.
    \item \textbf{Inconsistencies Handling} is divided into three sub-steps: the identification and imputation of statistical outliers (QE31), the detection and correction of typographical errors (QE32), and the identification and correction of logic errors (QE33). None of these sub-steps necessitates splitting the actions, as for these tasks, the processes of identifying the problem and resolving it occur simultaneously. These steps do not require the identification of a primary key. They only benefit from the PQE phase by targeting the right columns to deal with.
\end{itemize}
These steps collectively enhance data quality by systematically removing errors and inconsistencies, thereby preparing the dataset for accurate and reliable analysis.

It is noteworthy that in Figure \ref{fig:framework}, there is an arrow linking logic errors to the imputation of missing values. This is because, as we will see in Sections \ref{sec:Abs} and \ref{sec:LogicErr}, some specific types of missing values will be handled as logic errors.

\subsection{Pre-quality enhancement phase}
\subsubsection{Identification of the primary key in the data set}
\label{sec:primkeys}
Given a data set \( D \) consisting of \( n \) records \( \{ r_1, r_2, \ldots, r_n \} \), where each record \( r_i \) is a tuple \( (a_{i1}, a_{i2}, \ldots, a_{im}) \) of \( m \) attributes, a \textit{primary key} is a subset of the set of attributes (columns) $\{A_j\}_{j \in \llbracket 1, m \rrbracket}$ that uniquely identifies each record in the data set. Let \( P = \{ p_1, p_2, \ldots, p_k \} \subseteq \{A_1, A_2, \ldots, A_m\} \) be the set of $k$ fields corresponding to the primary key attributes and $P_{ind} \subseteq \{1, 2, \ldots, m\}$ the respective indices of each element of $P$. The primary key \( P \) must satisfy the following two properties:
\begin{enumerate}
    \item \textbf{Uniqueness} \mbox{}\\
    For any two distinct tuples (rows) \( r_i \) and \( r_j \) in the table where \( i \neq j \), their projections on the primary key attributes must be different, it thus, must hold that 
    $$
        r_i^P \neq r_j^P \quad \forall \, r_i, r_j \in D \text{ with } i \neq j,
    $$
    Here, \( r^P \) denotes the projection of tuple \( r \) onto the subset of attributes \( P \) which is 
    $$
        r_i^P = (a_{ij})_{j \in P_{ind}}.
    $$
    
    This ensures that no two rows can have the same value for \( P \), providing a unique identifier for each record.
    
    \item \textbf{Minimality} \mbox{}\\
    The set \( P \) must be minimal with respect to the uniqueness property. This means that no proper subset of \( P \) satisfies the uniqueness condition. If any attribute is removed from \( P \), the remaining attributes no longer uniquely identify every row in the table. It must therefore hold that if \( Q \subset P \) and \( Q \neq \emptyset \), then there exist at least two records \( r_i \) and \( r_j \) in \( D \) such that:
   \[
   r_i^Q = r_j^Q,
   \]
   where \( r_i^Q \) denotes the projection of record \( r_i \) onto the subset \( Q \).
\end{enumerate}

Primary keys in a database play a crucial role in ensuring data integrity and efficient data management. They are fundamental because they uniquely identify each record within a table, which is essential for maintaining the uniqueness and accuracy of the data stored. This unique identification facilitates efficient data retrieval, management, and manipulation processes, ensuring that each record can be accessed, updated, or related to other data in the database without ambiguity \citep{kohler2015,tranle2012,zhou2014}. They will be pivotal to finding duplicates and will shorten processing times when finding missing values.

The approach we introduce combines two strategies of keys identification. Pattern recognition and uniqueness analysis. The procedure is summarized as follow: 

\begin{enumerate}
    \item Theoretically, the set of attributes that constitute a primary key should never have missing values, particularly if the data is stored in a database software. However, since this framework is designed to accommodate any dataset, such an assumption cannot be made. Nevertheless, we can anticipate primary key candidates to exhibit only a few missing values. Therefore, we identify all fields with less than 5\% percent missing values.
    
    \item We then identify each column name that could potentially designate a primary key, as they typically contain terms such as \textit{ID}, \textit{CODE}, or \textit{KEY}. From there, two possibilities arise:
    \begin{enumerate}
        \item \textbf{Pattern recognition}\mbox{}\\
        We begin by attempting a quick-win solution. If this process has resulted in only one field remaining, then we have a good candidate for being the primary key. We conduct a set of sanity checks to ensure that the choice is justified. We verify that we do not have more than 5\% of duplicates (see below). Additionally, we ensure that the field is not a numerical valued field. While theoretically, a numerical valued field could serve as a primary key, it is rather unlikely due to format changes, uniqueness issues, etc. If all the sanity checks are successfully passed, we have identified our unique primary key.

        \item \textbf{Uniqueness analysis}\mbox{}\\
        If not, we proceed with candidates obtained from step 1 and conduct a uniqueness analysis for all combinations of variables, ranging from the smallest to the largest set. Similar to the fact that we cannot assume the fields comprising the primary keys will be devoid of missing values, we cannot assume they will lack duplicates. However, we can anticipate that only a few duplicates will occur. Again, we limit them to 5\%. Therefore, as soon as a combination of candidate attributes, with its number of duplicates falling below the threshold, is identified, it is considered as a primary key for the entire dataset. As fields labeled with the expressions \textit{ID}, \textit{CODE}, or \textit{KEY} are still likely candidates to form sets of primary keys, we initiate the process with them.    
    \end{enumerate}
\end{enumerate}

We can now analyze the complexity of this algorithm. It is clear that if we are in a case where the pattern recognition was not conclusive (step 2.(a)), the complexity is dominated by the uniqueness analysis step. Also, it is clear that the advantages obtained from the identification of the primary key, in terms of structuring the data set and accelerating other processes, might be outweighed by the resources deployed to obtain said primary key if that process is not controlled. For that reason, it is more computationally reasonable to limit the analysis of combinations analyzed for uniqueness. For instance, when considering only pairs, the worst-case complexity is $O(m^2  n  \log n)$, where $n$ is the number of records and $m$ is the number of attributes. Meanwhile, if the pattern recognition is successful, the time complexity is $O(m  n)$. In both cases, the space complexity is $O(m  n)$, primarily due to the storage requirements for intermediate results during uniqueness checks. In practice, however, as we will see in section \ref{sec:Resprimekeys}, the worst-case scenario is not reached as we start with the fields identified in the pattern recognition step. Indeed, if those fields can constitute the primary key, let $q$ be their number, then the complexity would be $O(q^2  n  \log n)$.

\begin{itemize}
    \item \textbf{Time complexity analysis}
    \begin{itemize}
        \item The initial step of identifying fields with less than 5\% missing values involves iterating through each field (column) and checking for missing values across all rows. This process results in a time complexity of $O(n  m)$, where $n$ represents the number of rows and $m$ represents the number of columns. This step is essential to filter out columns with excessive missing data, ensuring that potential primary key candidates are viable.
    
        \item Next, the process of identifying columns that could potentially designate a primary key by checking for specific terms such as "ID," "CODE," or "KEY" in their names has a time complexity of $O(m)$. This is because each column name is examined individually. This step narrows down the list of potential primary keys based on naming conventions, serving as a quick heuristic check. The pattern recognition step offers a quick-win solution where, if only one field remains, sanity checks are conducted to verify its suitability as a primary key. However, if the quick-win solution is not applicable, the algorithm proceeds to the more intensive uniqueness analysis.
    
        \item The uniqueness analysis step is the most computationally expensive part of the algorithm. It involves examining combinations of candidate attributes to ensure their uniqueness. If we consider combinations of size $s$, the number of such combinations is given by $\binom{m}{s}$, which is $O(m^s)$. The uniqueness check for each combination can be performed by sorting records, which has a $O(n  \log n)$ time complexity. Thus, the overall time complexity is $O(m^s) \times O(n  \log n) = O(m^s  n  \log n)$. In the case of pairs and focusing on candidate fields, the time complexity is $O(q^2  n  \log n)$.  
    \end{itemize}
    
    \item \textbf{Space complexity analysis}
    \begin{itemize}
        \item The space complexity for identifying fields with less than 5\% missing values is $O(m)$, as it requires storing a list of columns that meet the criteria. This step ensures that only relevant columns are considered for primary key candidacy without consuming excessive space.

        \item For identifying columns based on specific terms in their names, the space complexity remains $O(m)$. The list of potential primary key columns is stored, which is proportional to the number of columns in the dataset.

        \item The uniqueness analysis step has a space complexity of $O(m  n)$. This is because the algorithm needs to store combinations of columns and intermediate results for each uniqueness check. The space required for these combinations and results can grow significantly with an increasing number of columns, making this step the most space-intensive part of the algorithm.
    \end{itemize}
\end{itemize}

\subsubsection{Mapping processes to specific data fields}
\label{sec:Map}
As our work aims to enhance and measure data quality without relying on domain knowledge or knowledge of the current table's content, it is crucial to be able to precisely target each analysis and correction to maintain reasonable computer resource demands. A similar approach is taken by \cite{schelter2018automating}. While the authors do not specifically aim to make their work domain-knowledge-free, their software can propose a set of constraints to verify for each column in the absence of user-provided constraints. This resembles our problem, but according to the authors, their constraint suggestion process is designed to involve human intervention. While the automated process we introduce may be time-consuming, it ultimately saves time and computational resources when considering the entire process. The defined rules for each type of error considered in this work are as follows:

\begin{itemize}
    \item \textbf{Redundancies} \mbox{}\\
    Only the subset of attributes comprising the primary key will be analyzed for duplicates. This is one of the justifications for why it is important to begin by identifying it.
    \item \textbf{Absences} \mbox{}\\
    All fields will be analyzed for missing values, as they can affect any field. However, as we will see in Subsection \ref{sec:Abs}, not all fields' missing values will be imputed.
        
    \item \textbf{Statistical outliers} \mbox{}\\
    Only fields with numerical values will be considered. We also assert that without sufficient information on the field, accurately apprehending its statistical properties is challenging. Hence, we will not attempt to detect statistical outliers unless more than half of the values are available.

    \item \textbf{Typographical errors} \mbox{}\\
    The algorithm introduced in Subsection \ref{sec:Typos} functions for both real words and non-words. However, it is not intended for application to entries containing both numerical and text. For instance, the field \textit{fiBaseModel} includes values like \textit{ZX160}. Without domain knowledge, it is at best very difficult to ascertain whether another entry, such as \textit{ZX161}, is a typographical error or a legitimate model. For this treatment, we also exclude fields with less than half of the values available. Indeed, we argue again that with more than half of the data missing, any analysis or insights derived from this field are already significantly compromised. The presence of typographical errors in a minority of the data is less likely to have a substantial impact on the overall analysis compared to the large proportion of missing values.

    \item \textbf{Logic errors} \mbox{}\\
    These fields should contain strings, have less than 75\% of their values missing, and be represented by at least five different observations. Allowing up to 75\% missing values in these columns considers the unique nature of logic errors—errors that might not necessarily be about the absence of data but about data being present when it shouldn't be, or missing when it should be present. Therefore, when analyzing logic errors, it is not conservative enough to disqualify fields solely because they do not have enough data. Moreover, the requirement for at least five different observations ensures a diversity of data entries that can help identify inconsistencies and errors that a smaller number of observations might miss. By setting these parameters, we aim to enhance the reliability and validity of the corrections, focusing on fields where the scope for logic errors is significant and where corrections can substantially improve data quality.
    
\end{itemize}

\subsection{Quality enhancement phase}
\subsubsection{Redundancies handling}
\label{sec:Dupes}

\paragraph{Methodology}\mbox{}\\
Given our data set \( D \) consisting of \( n \) records \( \{ r_1, r_2, \ldots, r_n \} \), where each record \( r_i \) is a tuple \( (a_{i1}, a_{i2}, \ldots, a_{im}) \) of \( m \) attributes, we aim to identify duplicates based on the primary key attributes. We recall that in Subsection \ref{sec:primkeys} we defined \( P \subseteq \{A_1, A_2, \ldots, A_k\} \), the set of attributes defining the primary key. The primary key \( P \) must satisfy the properties of uniqueness and minimality, ensuring that each record in \( D \) is uniquely identifiable by \( P \). Two records \( r_i \) and \( r_j \) are considered duplicates if their projections on the primary key attributes are identical, i.e., \( r_i^P = r_j^P \). Therefore, the set of duplicates \( \mathcal{D}\mathcal{U}\mathcal{P} \) is defined as:
$$
\mathcal{D}\mathcal{U}\mathcal{P} = \{ (r_i, r_j) \in D \times D \mid i \neq j \text{ and } r_i^P = r_j^P \}.
$$ 
To generalize this concept to sets of more than two records, we define a set \( S \subseteq D \) as duplicates if all records in \( S \) have identical projections on the primary key attributes which is: $\forall \, r_i, r_j \in S, \quad r_i^P = r_j^P$. We hence have, the more generalized expression:
$$
\mathcal{D}\mathcal{U}\mathcal{P} = \{ S \subseteq D \mid \forall \, r_i, r_j \in S, \, r_i^P = r_j^P \text{ and } |S| > 1 \}.
$$

To identify all duplicate sets in \( D \), we follow these steps:

\begin{enumerate}
\item \textbf{Projection} \mbox{}\\ 
We start by computing the projection of each record onto the primary key attributes:
   \[
   \{ r_1^P, r_2^P, \ldots, r_n^P \}.
   \]

\item \textbf{Grouping} \mbox{}\\ 
Then, we group records by their primary key projections. Let \( G \) be the collection of $L$ groups, where $\forall \; l \in \llbracket 1, L \rrbracket$ each group  $G_l$,  contains records with the same primary key projection:
   $$
   G = \{ G_1, G_2, \ldots, G_L \},
   $$
   where \( G_l = \{ r_i \in D \mid r_i^P = v_l \} \) for some unique primary key projection \( v_l \).

\item \textbf{Identifying duplicate sets} \mbox{}\\ 
We now can identify all groups \( G_l \) in \( G \) that contain more than one record. These groups represent sets of duplicate records:
   $$
   \mathcal{D}\mathcal{U}\mathcal{P} = \{ G_l \mid |G_l| > 1 \}.
   $$

\item \textbf{Dropping duplicates} \mbox{} \\
Finally, for all subsets $G_l$, we drop all records but one, we typically keep the first instance.
\end{enumerate}

\paragraph{Complexity analysis}\mbox{}\\
This generalized approach ensures that any set of records with identical primary key attribute values is identified as duplicates\footnote{In practice, in our implementation, we used the \textit{drop_duplicated} method from Python's pandas library \citep{pandas}}. Let's analyze its time and space complexity. It can be analyzed in terms of the number of records $n$ in the data set $D$ and the number of attributes $k$.

\begin{itemize}
    \item \textbf{Time complexity} 
    \begin{itemize}
        \item First, we consider the \textbf{projection} step, where we compute the projection of each record onto the primary key attributes. This involves iterating over all \( n \) records and projecting each record onto the \( k \) primary key attributes. If \( k \) is relatively small compared to \( n \), the projection step has a time complexity of \( O(n) \), otherwise if \( k \) is large, the time required for each projection increases. Specifically, the total time complexity of the projection phase is \( O(n k) \). 

        \item Next, in the \textbf{grouping} step, we group records by their primary key projections. This step involves inserting \( n \) projected records into a suitable data structure for grouping (typically a dictionary or a hash table). Inserting each projected typically has an average-case time complexity of \( O(1) \). Therefore, the grouping step has an average-case time complexity of \( O(n) \). 

       \item  Then, in the \textbf{duplicate identification} step, we identify all groups that contain more than one record. This step involves iterating over the groups formed in the previous step. In the worst case, there could be up to \( n \) groups (if all records have unique primary key projections). Checking the size of each group and collecting groups with more than one record has a time complexity of \( O(n) \). 
    
        \item Finally, for each group $G_l$ that contains more than one record, we keep only the first record and drop the rest. Iterating over all groups takes $O(L)$ time. Again, in the worst case, we have $n$ groups. For each group $G_l$, dropping the records but the first one takes $O(|G_l|)$ where $|G_l|$ is the number of records in the group $l$. The sum of all records in the different groups being $n$. The time complexity of this step is therefore $O(n)$.
    \end{itemize}
        Therefore, the overall time complexity is $O(n  k + n + n + n) = O(n  k)$  when $k$ is large. This complexity is reduced to $O(n)$ when $n >> k$.
    
    \item \textbf{Space complexity}\mbox{} \\ 
        The space complexity of the algorithm is also important to consider. The projections of the records require $O(n)  k$ space for a large $k$ and $O(n)$ otherwise. The hash table used for grouping requires $O(n)$ space. Collecting the duplicate sets requires $O(n)$ space in the worst case (if all records are duplicates). Dropping the duplicates does not require any additional storage. Therefore, the overall space complexity is $O(n)  k$.

    \end{itemize}

This analysis indicates that if the number of attributes $k$ defining the primary key PP is reduced, both time and space complexity become $O(n)$. However, for a large value of $k$, the time and space complexities of the algorithm become $O(n  k)$, which could impact performance for large datasets. The typical scenario occurs when no previous work has been done to identify a primary key, resulting in $k$ being exactly equal to the number of fields in the database (53 in our case). This underscores the importance of the pre-quality enhancement phase, as it not only allows for better structuring of the dataset but also reduces the computational cost of the algorithms in terms of both time and space complexities.

\subsubsection{Missing values handling}
\label{sec:Abs}
As we have already stated, identifying missing values is not a tedious task in itself. The real challenge lies in distinguishing justifiably missing values from others and correcting only these values. Without domain knowledge, this task becomes even more difficult. In this section, we decide to implement a rather simple two-step solution.

\begin{itemize}
    \item We identify fields that have missing values which might be unjustified. These fields fall into two categories: first, the attributes of the primary key; and second, the attributes that don't have many missing values. We consistently maintain the same 95\% threshold used for primary key candidates (see Subsection \ref{sec:primkeys}). Any field with more than 95\% missing values will be considered to have justified missing values, and we will not attempt to impute these missing values. In an industrialized implementation, such fields would be flagged by \textit{warnings}.

    \item If the field being analyzed is part of the primary key, the missing value is automatically deemed abnormal. Since the projection of any record on the primary key cannot be duplicated, we cannot use the distribution of values in the same field for this case (even if the primary key might be comprised of multiple fields, this approach is not conservative enough). Therefore, we automatically generate a placeholder character that will act as the record for the remainder of the process.
    Then if the field being analyzed is not part of the primary key, instead of attempting to determine whether a missing value is justified or not, we assume that any observation, including missing values, could be valid and there are two possibilities: 

    \begin{itemize}
        \item If the field is a character-valued field, no imputation is made, and we assume that this missing value is justified. In the remainder of the process, specifically in Subsection \ref{sec:LogicErr}, this will be checked as a logical error. For instance, if the \textit{Enclosure} field in one row is empty, the algorithm introduced later will determine if it is likely for the model description (field \textit{fiModelDesc}) in the considered row to have an empty Enclosure - without using domain knowledge -.

        \item If the field contains numerical values, we apply a simple and classical method by replacing the missing values using linear interpolation with the two closest present values (one with a lower index and one with a higher index). In this work, where we do not have detailed information about the dataset, it is not advisable to replace missing values with a statistic (typically the average or the median) of the observations of the specific field. Indeed, the dataset could, for instance, contain time series with seasonality. In that case, the average might not be suited. 
    \end{itemize}
\end{itemize}
 
\subsubsection{Processing statistical outliers}
\label{sec:outliers}
Let us now focus on identifying and correcting statistical outliers. 
We recall that we had defined a dataset $D$ as a finite collection of $n$ records $\{r_i\}_{i \in \llbracket 1, n \rrbracket}$ with each $r_i$ having $k$ attributes, i.e., $r_i = (a_{i1}, \dots, a_{ik})$. We are interested in analyzing the outliers for a given attribute. That is to say, we want to determine whether, for a given attribute $j$ of record $r_i$, the observation $a_{ij}$ is homogeneous with the rest of the observations $\{a_{lj}\}_{l \in \llbracket 1, n \rrbracket \setminus \{i\}}$. Of course, depending on the meaning given to \textit{homogeneous}, the identification and handling of statistical outliers may vary.

As we have discussed in Section \ref{sec:Intro}, several techniques have been developed to identify outliers. Among them, many use classical results on interquantile intervals. This is notably the case with \citep{CITE_12}, which decides that valid states $a_{ij}$ are those that verify $a_{ij} \in [Q_1 - 1.5(Q_3-Q_1),\;Q_3 + 1.5(Q_3-Q_1)]$ where $Q_1$ and $Q_3$ define respectively the first and third quartiles of the distribution of attribute $j$'s observations $\{a_{lj}\}_{l \in\llbracket 1, n \rrbracket}$. Another common method (see for instance \cite{CITE_24}) involves invalidating all values that lie outside an interval based on the standard deviation. In other words, $a_{ij}$ is invalidated as soon as $a_{ij} \notin [-\alpha\sigma + \mu,\; \alpha\sigma + \mu],\; \alpha \in \mathbb{R}^{*}_{+}$. $\mu$ designates the mean of the distribution and $\sigma$ its standard deviation. The most common practice seems to be to choose $\alpha = 3$ (\cite{CITE_25} for example).

There are several limits to these methods:

\begin{itemize}
    \item Firstly, they assume that the distribution of observations is normal \citep{CITE_26}.
    \item Secondly, the mean and standard deviation are strongly influenced by the outliers we want to detect \citep{CITE_26}, leading to distorted measures of central tendency and dispersion.
    \item Finally, as \cite{CITE_27} points out, this method has very little chance of detecting outliers in small samples.
\end{itemize}
Therefore, applying these methods can lead to incorrect identification of outliers. In skewed distributions, one might either miss true outliers or falsely identify regular data points as outliers. For multimodal distributions, these methods might fail to recognize outliers within each mode. Overall, there is a non-negligible chance of falsely identifying normal data points when the assumptions required by these methods are not met. Despite these disadvantages, there are good reasons to continue using these methods. They are very simple to use and interpret, and when the amount of data used is large, the sample can approach a Gaussian distribution.

\begin{figure}[H]
    \centering
    \includegraphics[width=7.5cm]{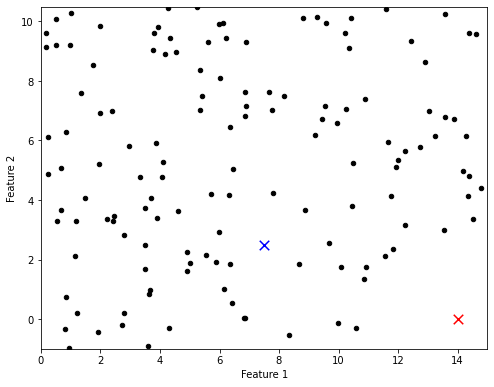}
    \caption{Dummy data set, the red and blue points are the two points we will focus our illustration on.}
    \label{fig:if1}
\end{figure}

\begin{figure}[H]
    \centering
    \begin{minipage}{0.5\textwidth}
        \centering
        \includegraphics[width=7.5cm]{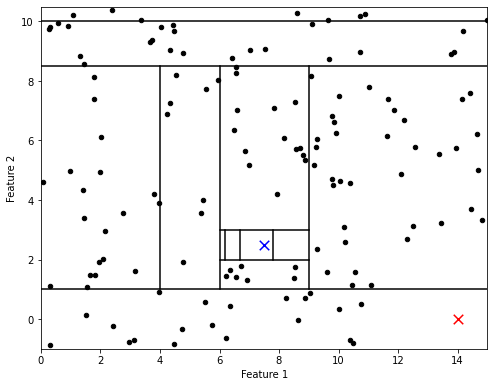} 
        \caption{Isolation tree on the blue point}
        \label{fig:if2}
    \end{minipage}\hfill
    \begin{minipage}{0.5\textwidth}
        \centering
        \includegraphics[width=7.5cm]{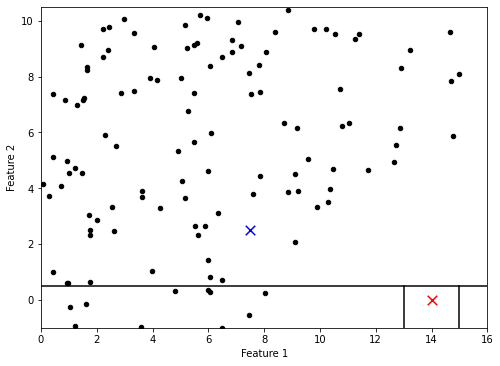} 
        \caption{Isolation tree on the red point}
        \label{fig:if3}
    \end{minipage}
\end{figure}

\paragraph{Isolation Forest}\mbox{}\\
As others have done in the literature (see section \ref{sec:Intro}), we want to leverage machine learning's ability to overcome the above limitations. A widely utilized and highly effective algorithm for anomaly detection that takes a novel approach to identifying outliers is Isolation Forest (IF) \citep{CITE_29}.  Unlike traditional methods that first establish what constitutes normal data and then identify deviations from it, Isolation Forest focuses directly on isolating anomalies. The core idea behind this algorithm is that anomalies are easier to isolate than normal points because they are \textit{few and different}. By recursively partitioning the data set, the algorithm creates a series of decision trees designed not to group similar data points but to isolate individual points. Anomalies, due to their rarity and distinctiveness, require fewer partitions to be isolated compared to normal data points. This process makes anomalies stand out more clearly and be detected more efficiently. The process and efficiency of Isolation Forest in isolating outliers is visually depicted in Figures \ref{fig:if1}, \ref{fig:if2} and \ref{fig:if3}, where the partitions highlight the isolation of anomalies from the rest of the data set. This method is particularly advantageous in large datasets and real-time anomaly detection scenarios due to its linear time complexity and scalability. As this method seems less common than other used in this work, we give a more detailed explanation of the algorithm's functioning which can be described by the following steps: 

\begin{enumerate}
    \item \textbf{Subsampling:}
    \begin{itemize}
        \item The algorithm randomly selects a subset of the data points from the dataset. This step helps make the algorithm scalable and reduces computational complexity.
    \end{itemize}

    \item \textbf{Tree Construction:}
    \begin{itemize}
        \item Recursive Partitioning: For each selected subset, the algorithm recursively partitions the data points by randomly selecting a feature and then randomly selecting a split value for that feature.

        \item Node Creation: Each partition creates a node in the tree. The process continues until each data point is isolated in its own leaf node, the maximum tree height is reached, or the node contains a single data point.
    \end{itemize}

    \item \textbf{Forest Building:}
    \begin{itemize}
        \item The tree construction process is repeated multiple times to build a collection of isolation trees, forming the Isolation Forest.
    \end{itemize}

    \item \textbf{Path Length Calculation:}
    \begin{itemize}
        \item For each data point in the dataset, the algorithm computes the average path length from the root node to the terminating node (leaf) across all the trees in the forest. The path length is the number of edges traversed from the root to the leaf.
    \end{itemize}

    \item \textbf{Anomaly Score Computation:}
    \begin{itemize}
        \item The anomaly score for each data point is calculated based on the average path length. The score is defined such that shorter average path lengths correspond to outliers (since outliers tend to be isolated quickly), while longer average path lengths correspond to normal data points.
    \end{itemize}

    \item \textbf{Anomaly Thresholding:}
    \begin{itemize}
        \item The algorithm determines a threshold for the anomaly score. Data points with anomaly scores above this threshold are classified as outliers, while those below the threshold are considered normal.
    \end{itemize}
\end{enumerate}

For sake of illustration, let us consider a dummy data set, with two features (attributes) \textit{Feature 1} and \textit{Feature 2} represented by figure \ref{fig:if1}

Now let us consider two isolation trees focusing respectively one the blue and the red point.

\begin{figure}[H]
    \centering
    \includegraphics[width=7.5cm]{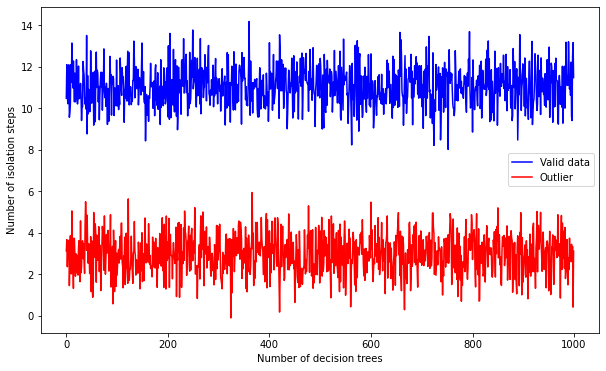}
    \caption{Number of steps for a valid data and a statistical outlier}
    \label{fig:if4}
\end{figure}

Figures \ref{fig:if2} and \ref{fig:if3} illustrate the isolation process for two observations: one that seems to be valid (the blue cross) and one that seems to be an outlier (the red cross). We can easily understand how the isolation forest algorithm works. The blue cross is isolated after many steps while only three steps are enough to isolate the red cross\footnote{Note that these graphics are for the sake of illustration; the dataset and tree parameterization are chosen to perfectly and only match the isolation of the red and blue crosses. In practice, the isolation analysis is performed for all observations simultaneously}. This indicates that the red cross is an outlier. This process is repeated over a large number of random trees (leading to the forest). Then, the number of steps (branches) before the splitting are averaged (step 4 above), and a comparative analysis is performed with the threshold (steps 5 and 6 above). Figure \ref{fig:if4} shows the average number of steps for different sizes of forests.

This algorithm has proven its efficiency, particularly on large datasets. However, this is precisely where the limitation lies in our context. As this method is more interesting in high dimensions, it naturally loses its explainability, particularly in point 1.c of the definition (error identification). In fact, if the dataset contains only one or two columns, the algorithm loses its efficiency (as can be intuited from Figures \ref{fig:if2} and \ref{fig:if3} above). On the other hand, if it uses all the variables or a subset of them, it becomes impossible to know which column really poses a problem. We therefore choose not to use this powerful technique in high dimensions. Instead, we propose to combine statistical methods using standard deviation and IF in one dimension.

We work with $Z$ the $Z$-score of observation $a_{ij}$, $Z = \frac{a_{ij}-\mu_j}{\sigma_j}$. Where $\mu_j$ and $\sigma_j$ designate the unbiased estimators of  respectively the expectation and the standard deviation of the distribution of attribute $j$'s observations $\{a_{lj}\}_{l \in\llbracket 1, n \rrbracket}$. Let $\varphi_{outlier}$ be the function defining whether the input $X$ is an outlier, returning 1 in that case and 0 otherwise. Let us also assume that our IF is a function that returns 1 when given an outlier and 0 otherwise. 
\begin{align*}
  \varphi_{\text{outlier}} \colon \mathbb{R} &\to \{0,\ 1\}\\
  Z &\mapsto \varphi_{\text{outlier}}(Z).
\end{align*}
\begin{equation}
\varphi_{\text{outlier}}(Z) = \left\{
    \begin{array}{ll}
        f_1(Z,\;\beta_1,\;\beta_2) & \mbox{if}\;|\mu_3|<\alpha_s\;\text{and}\;|\mu_4|<\alpha_k \\
        f_2(Z,\;\beta_1,\;\beta_2,\;\gamma) & \mbox{otherwise},
    \end{array}
\right.
\end{equation}
where:
\begin{equation}
f_1(Z,\;\beta_1,\;\beta_2) = \left\{
    \begin{array}{ll}
        1 & \mbox{if} \ Z \notin ]-\beta_1,\;\beta_2[ \\
        0 & \mbox{otherwise}
    \end{array}
\right.
\end{equation}
and
\begin{equation}
f_2(Z,\;\beta_1,\;\beta_2,\;\gamma) = f_1(Z,\;\gamma\beta_1,\;\gamma\beta_2) \times IF(Z),
\end{equation}
where

\begin{equation}
IF(Z) = \left\{
    \begin{array}{ll}
        1 & \mbox{if the Isolation Forest algorithm detects an outlier}  \\
        0 & \mbox{otherwise}.
    \end{array}
\right.
\end{equation}

We define the terms \footnote{The large number of parameters may seem like an obstacle to easy automation, but in reality, these parameters are quite simple to choose and correspond to fairly intuitive and flexible choices based on the desired tolerance.}:

\begin{itemize}
    \item $\mu_3$ and $\mu_4$ respectively define the $3^{rd}$ and $4^{th}$ standardized moment, skewness and kurtosis.
   
    \item $\alpha_s$ and $\alpha_k$ are two strictly positive moment acceptance thresholds. We take 6 and 30 respectively.
    \item $\beta_1$, $\beta_2$ are two strictly positive acceptance thresholds for the standard deviation-based tolerance interval. We take 3 for both coefficients.
    \item Finally, $\gamma$ is a parameter to widen the tolerance interval on the value of Z, we choose 2.
\end{itemize}

The idea behind this algorithm is to suggest that if we have enough data with a distribution that is not too distorted, with reasonable kurtosis and skewness, the Gaussian approximation is not a significant assumption. In this case, we can apply the standard deviation rule directly. If, on the opposite, this is not the case, we apply the same rule but with a tolerance interval $\gamma$ times larger (e.g., twice as large). Since we are losing reliability, we seek a \textit{second opinion} using the IF algorithm and consider as outliers only those entries where both techniques agree on the invalidity of the data. The main advantage of this method is that we benefit from the standard deviation's ability to discriminate values that are too high or too low, while the IF not only looks at extreme values but also identifies outliers as values that are isolated from the remaining dataset. Indeed, this algorithm will not identify an extreme value as an outlier if it has other extreme values in its neighborhood.

Figures \ref{fig:price} and \ref{fig:yearmade} below illustrate the Kernel Density Estimation (KDE) of two attributes: \textit{Saleprice} and \textit{YearMade}. They help easily understanding the algorithm's process. For the attribute \textit{YearMade}, the data is centered around the year 2000, with some noise around the mode of the distribution. We also observe that the support of the density reaches negative values as well as very high values, indicating the presence of invalid values. In this case, due to the relatively correct appearance of the density, we do not need the computational power required for the IF algorithm. We can be optimistic in our ability to successfully identify outliers using only the standard deviation rule, that is to say, $\varphi_{outlier}(Z) = f_1(Z, \, \dots)$. On the other hand, looking at the attribute \textit{Saleprice}'s density, we observe that the rightmost part of the density has high values. Using only the standard deviation rule, these values would probably be wrongfully discarded. Therefore, the algorithm widens the tolerance of the standard deviation rule to ensure that we focus on extreme values, and then the IF identifies isolated extreme values, ensuring that a price is not discarded for being high but for being in a high region where no other product has a close price. That is to say, $\varphi_{outlier}(Z) = f_2(Z, \,\dots)$. If we had domain knowledge, the solution we would apply would be very close to the one our algorithm applies. Indeed, we would have known that the prices of the products will generally be reasonably close, but for some specific products, the prices could be very high (hence the behavior of the distribution). 

\begin{figure}[H]
    \centering
    \begin{minipage}{0.5\textwidth}
        \centering
        \includegraphics[height=7cm, width=1\textwidth]{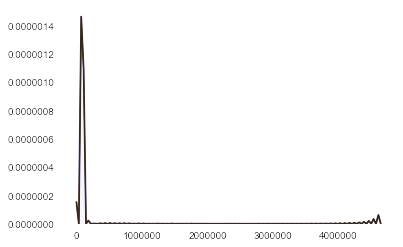} 
        \caption{KDE of attribute \emph{Saleprice}}
        \label{fig:price}
    \end{minipage}\hfill
    \begin{minipage}{0.5\textwidth}
        \centering
        \includegraphics[height=7cm, width=1\textwidth]{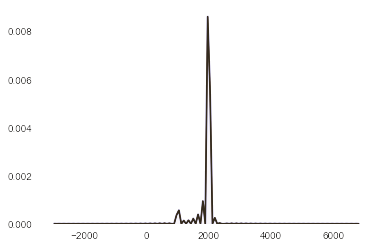} 
        \caption{KDE of attribute \emph{YearMade}}
        \label{fig:yearmade}
    \end{minipage}
\end{figure}

Once outliers have been rigorously identified as invalid, they are treated like any other numerical missing value. They are therefore imputed using linear interpolation, as described in Subsection \ref{sec:Abs}.

\paragraph{On the choice of $\alpha_s$ and $\alpha_k$}\mbox{}\\
The choice of the tolerance parameters for the skewness and the kurtosis, respectively $\alpha_s$ and $\alpha_k$, is important. One must keep in mind that the framework should work without domain knowledge and on any given dataset. Therefore, the values chosen must provide a practical, flexible, and robust framework for handling real-world data distributions, which often deviate from normality. We chose to select our parameters based on a known skewed and heavy-tailed distribution. We used a chi-squared distribution with 1 degree of freedom. The skewness threshold of 6 is approximately double the skewness of that distribution, indicating a substantial but reasonable allowance for asymmetry. The kurtosis threshold of 30, being double the kurtosis of the reference distribution, suggests tolerance for heavy-tailed distributions. These thresholds enable the primary outlier detection mechanism (based on standard deviation) to be broadly applicable, ensuring minor deviations in distribution shape do not overly influence outlier detection. This approach and the chosen values balance sensitivity and specificity, reducing false positives. The high thresholds ensure robustness, making the algorithm adaptable to various scenarios without frequent false alarms, while still being stringent in cases of extreme skewness or kurtosis. These parameters can be tuned depending on one's tolerance to false positives.

\paragraph{Complexity analysis}\mbox{}\\
The complexity of this algorithm is dominated by the IF component. A thorough analysis on its time and space complexity is conducted by \citep{CITE_29}. Their work demonstrates a time complexity of the IF algorithm in $O(t\psi \rm{log} \psi+nt\rm{log}\psi)$, and a bounded memory requirement that grows linearly with $n$. In these expressions, $t$ designates the number of trees, $\phi$ is the sub-sampling size, and $n$ is the number of records.

\subsubsection{Processing Typographical Errors}
\label{sec:Typos}
The correction of typographical errors (TPOs) is a very important issue in the preparation of data processing. Spell checking stands as the crucial process of identifying and suggesting corrections for words that are inaccurately spelled. Essentially, classic spell checkers serve as computational tools leveraging a dictionary of words to execute this task. The efficacy of such a checker relies on the extent of its dictionary. A larger repository of words typically results in an increased ability to detect errors. However, due to their reliance on standard dictionaries, they are inadequate in capturing mistakes such as proper nouns, specialized terms pertinent to particular domains, acronyms, and other specialized terminologies. These types of words are called non-words in the literature \citep{nonword}. They represent a notable contrast to conventional words, commonly referred to as real words. This is well explained in \cite{googletypo}. As we have already mentioned in Section \ref{sec:Intro}, on one hand, much of the literature requires the use of external dictionaries, and on the other hand, many algorithms in the literature use ML/DL techniques to handle domain knowledge, particularly to correct non-words. We introduce in what follows a procedure leveraging ML, efficient for both real and non-words without any use of domain knowledge. Let us start by introducing key concepts that are pivotal for our algorithm.

\paragraph{Damerau-Levenshtein Distance}\mbox{}\\
Consider two distinct character strings $A$ and $B$ and let us consider the task of finding the minimum number of operations required to transform $B$ into $A$.

When only substitutions are allowed, the minimum number of operations is given by the Hamming distance \citep{CITE_35}, or when only transpositions are allowed, the Jaro distance \citep{CITE_36}. The Levenshtein distance \citep{CITE_32} is the length of the shortest sequence when substitutions, insertions, and deletions are allowed. This distance is used in the longest common subsequence problem \citep{CITE_34}. In this paper, we will use the Damerau-Levenshtein distance \citep{CITE_32, CITE_33}, which is optimal when four operations are possible (substitution, insertion, deletion, and transposition of two consecutive characters). The Damerau-Levenshtein distance (DLD) for the letters number $i$ and $j$ of words $A$ and $B$ is recursively obtained through $d(.,.)$ below \citep{CITE_37}:
\begin{equation}
d_{A,B}(i,j) = \min 
    \begin{cases}
        0 & \text{if } i = j = 0 \\
        d_{A,B}(i-1,j) + 1 & \text{if } i > 0 \\
        d_{A,B}(i,j-1) + 1 & \text{if } j > 0 \\
        d_{A,B}(i-1,j-1) + 1_{(A_i \neq B_j)} &  \text{if } i, j > 0\\
        d_{A,B}(i-2,j-2) + 1 & \text{if } i, j > 1 \text{ and } A_i = B_{j-1} \text{ and } A_{i-1} = B_j.\\
    \end{cases}
\end{equation}

For example, since it only takes one transposition to go from \emph{France} to \emph{Fra\underline{cn}e}:
\begin{equation*}
    DLD(France,\;Fracne) = 1.
\end{equation*}

In practice, we will not use the DLD itself, but rather a score that increases the closer two words are to each other. Indeed, a distance of 2 between two 4-letters words, for example, and a distance of 2 between two 15-letters words should not be interpreted in the same way. The Damerau-Levenshtein Score (DLS) is given as a function of the DLD between $A$ and $B$, $DLD(A, B)$ and the maximum possible distance between two words of the lengths of $A$ and $B$, noted $DMAX_{A,\;B}$:
\begin{equation}
    DLS(A, B) = \frac{DMAX_{A,\;B} - DLD(A, B)}{DMAX_{A,\;B}}.    
\end{equation}

Considering the previous example, note that the maximal Damerau-Levenshtein distance between two words of length 6 is 6. Therefore, the score between \emph{France} and \emph{Fra\underline{cn}e} is given by:
\begin{equation*}
    DLS(France,\;Fracne) = \frac{6 - 1}{6} \approx 0.8334.
\end{equation*}

When considering now two words of length 3, for example \textit{Pau} and \textit{P\underline{o}u}, with still a DLD of 1, the score is lower than for the previous example:
\begin{equation*}
    DLS(Pau,\;Pou) = \frac{3 - 1}{3} \approx 0.667.
\end{equation*}

\paragraph{Detecting typographical errors}\mbox{}\\
To detect all typographical errors in a database field, the ideal solution would be to calculate the DLS of each entry with all the others. This would give us a matrix of DLSs, and we would then group together the entries with low DLSs, as this would indicate that they were spelled very similarly and were therefore the same word spelled in different ways. To find the true value and thus correct the TPOs, it would be logical to take the entry that has been used the most often. Unfortunately, this ideal solution is not feasible due to its computational cost. Indeed, even when only considering the step involved in building the DLS matrix, it would require $O(n^2)$ Damerau-Levenshtein function calls, which is not feasible for a table with hundreds of thousands or even millions of entries. Nevertheless, we have decided to keep the essence of this method but attempt to reduce the size of the problem. To achieve this, we will sort the list of observations alphabetically before calculating the consecutive DLSs. The idea is that by sorting them in alphabetical order, we will increase our chances of grouping similar entries together and forming part of the aforementioned groups. 

\begin{figure}[H]
    \centering
    \includegraphics[height=8.cm, width = 16.4cm, angle=0]{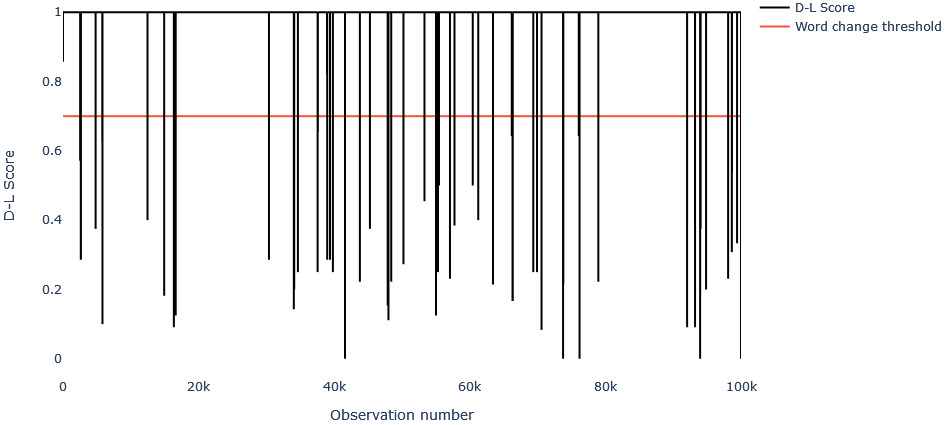}
    \label{fig:dljumps}
    \caption{Damerau-Levenshtein score jumps for variable \textit{State}}
\end{figure}

To identify the groups, we examine the curve of consecutive DLSs. As soon as a jump is observed on the DLS curve, we have a priori reached a different word. The threshold beyond which we admit to having changed words is set at 0.7. Figure \ref{fig:dljumps} illustrates the curve of DLSs of words alphabetically ordered as well as the threshold. The words correspond to the entries of the field \textit{State}.

Having identified all the groups based on the DLS jumps, we can now represent each of these groups by their \textit{dominant} element (the one with the most occurrences in the dataset). It is then necessary to retrieve any elements of a group we may have missed. For example, \textit{France} and \textit{\underline{G}rance} are very close in DLS terms, but when sorted alphabetically, they will not be in the same group. This is the step where we use ML, specifically unsupervised learning via clustering techniques. The clustering algorithm seeks to group together all words with similar DLS vectors (a row of the DLS matrix). 

The algorithm we decided to use is hierarchical agglomerative clustering \citep{CITE_38}. In a different context from that of this document, where the use of a priori knowledge of the database could be exploited, the user could communicate the correct value (which could correspond, for instance, to the number of states where sales have been concluded). However, in our context, we are obliged to determine the number of clusters. To do this, we use the gap statistic method \citep{CITE_39}, which proceeds by comparing the intra-cluster dispersion with that which would have been obtained in an empty set.

At this point, all similar elements have been grouped together. However, an additional step is necessary: correcting wrongfully grouped elements. While this situation is unlikely, it can occur when dealing with long and similar words. For example, \textit{North Dakota} and \textit{South Dakota} share ten out of twelve characters, resulting in a DLS of 0.8334, which exceeds the 0.7 threshold. 

The introduced solution is quite straightforward. Within each group, we identify elements that differ from the dominant element (the one with the most occurrences) but still have high frequencies. These entries are candidates for being incorrectly attributed to a group. Depending on the nature of the field being analyzed, one of two solutions will be applied. If the words being tested are dictionary entries, a simple test is to determine if the elements differing from the dominant element with high occurrences are valid dictionary entries. If they are, they are identified as valid entries, and a flag (typically a \textit{warning} in programming languages) is raised to prompt the user to check theses specific corrections after the whole process. On the other hand, if the words are not dictionary entries but non-words, no correction is made due to insufficient information to confidently discard the previously rigorously obtained result. Again, a flag is raised for the user to review. 

\paragraph{Remarks}\mbox{}
\begin{itemize}
    \item It is important to note that this process, especially checking if the words are real or non-words, is not expensive since we only need to consider the unique dominant entries, whose number is significantly lower than the number of observations.
    
    \item The aim of this work is to construct a fully automated process to measure and enhance the quality of a data set. Flags are raised solely to inform the user of corrections that were or were not conducted in cases of potential ambiguity. No action is required from the user, ensuring that the process remains fully automated.

    \item It is important to underline the significance of the choice of the 0.7 threshold to detect DLS jumps. This is a hyper-parameter of the algorithm. The higher the threshold, the fewer the number of mistakenly grouped words, but as a trade-off, the number of elements to include in the clustering analysis increases. On the other hand, the lower the threshold, the more mistakenly grouped words, and therefore, more corrections are required after clustering. To strike the right balance between accuracy and computational performance, it is necessary to carefully choose the threshold. A higher or lower threshold could yield better or worse results depending on the data set. Practitioners should note that it is better to use a higher threshold (typically close to 0.7) and recover words that were not initially grouped through clustering rather than grouping words incorrectly from the start.

\end{itemize}

The TPO detection algorithm described can be summarized as follows:

\begin{enumerate}
    \item The list of words requiring TPO detection is sorted alphabetically, and the consecutive DLSs are calculated.
    \item Using DLS jumps, we distinguish the initial similarity groups. Each group contains all the elements from the current peak (included) to the next (excluded). Each group is then represented by its dominant element (the element with the highest frequency).
    \item Dominant elements that seem to represent the same word are regrouped. This is achieved by calculating the matrix of DLSs on all dominant elements, then clustering this matrix to regroup similar words. Dominant elements that are in the same clusters are considered to represent the same word. Their initial groups are merged. The new dominant element is the one among them with the highest occurrence.
    \item A final verification is made to ensure that no words are wrongfully flagged as a TPO. In each cluster, we identify non-dominant elements with high frequencies (for instance, members of a cluster that have more than half the frequency of the dominant element), these have potentially been wrongfully identified as typographical errors.
    \begin{enumerate}
        \item If the field being analyzed contains real words that are listed in the dictionary, we check whether the potentially wrongfully identified words are valid dictionary entries, if so, they are considered as valid words and not TPOs.
        \item Otherwise, we do not make any modifications to the previously constructed groups.
    \end{enumerate}
    \item At the end of the process, we have as many groups as valid words (real words or non-words, depending on the field content). Each group is represented by one and unique dominant element, and each element in the group written differently will be corrected to the dominant element.
\end{enumerate}

\paragraph{Complexity analysis}\mbox{}\\
Let us analyze the time and space complexity of this algorithm. We recall that $n$ is the number of records in the dataset, thus it is the number of words in a given field. Let us define $g$ as the number of dominant elements, $g<<n$. Let $l_{\rm{avg}}$ be the average length of the words being corrected. $l_{\rm{avg}}$ is a negligible value compared to $n$.

\begin{itemize}
    \item \textbf{Time complexity} 
    \begin{itemize}
        \item \textbf{Sorting the list of words alphabetically} takes $O(n\log n)$ time.
        \item The DLS calculation between two words has an average complexity of $O(l_{\rm{avg}}^2)$. \textbf{Calculating DLS between consecutive words} in the sorted list has an average complexity of $O(n l_{\rm{avg}}^2)$. We can reasonably state that $l_{\rm{avg}}^2 << n$. Which leads to a time complexity of $O(n)$ for this step.
        
        \item \textbf{Distinguishing initial similarity groups and determining dominant elements} involves scanning through the list of words once to form groups and determine the dominant element within each group. This is $O(n)$ since it involves a linear pass through the list.
        
        \item To \textbf{regroup dominant elements by clustering DLS matrix}, we start by constructing the DLS matrix for $g$ dominant element which involves $g(g-1)/2$ DLS calculations, each taking $O(l_{\rm{avg}}^2)$, resulting in $O(g^2  l_{\rm{avg}}^2)$. Here we can not make the assumption $l_{\rm{avg}}^2<<g^2$. Clustering the matrix using hierarchical clustering has a complexity of $O(g^3)$.
        
        \item The last step of the algorithm consists in \textbf{verifying flagged TPOs}. It requires identifying non-dominant elements with high frequencies by scanning through the whole list, resulting in a complexity of $O(n)$.
        Then, checking against a dictionary (assuming the dictionary check is $O(1)$) involves scanning each flagged word, taking $O(n)$ in the worst case. 
        
        \item \textbf{Constructing the final groups and correcting words}:
            This step involves scanning through the list and making necessary corrections and therefore has a time complexity of $O(n)$.
    \end{itemize}
    Let us now aggregate all complexities. Based on the analysis conducted for each of the previous steps, the overall complexity is $O\left( n \log n + n + g^2  l_{\rm{avg}}^2 + g^3  \right)$, which is equivalent to $O\left( n \log n + g^2  l_{\rm{avg}}^2 + g^3  \right)$. This complexity analysis again demonstrates how beneficial it is, from a computational point of view, to reduce the clustering process dimension by sorting the elements and only taking dominant elements. When $g$ is very small, the overall complexity becomes $O(n\log n)$ whereas the algorithm without the reduction through $g^3$ would have an overall complexity of $O(n^3)$. This low complexity, justifies the very good computational performance obtained as we will show in Section \ref{sec:Results}.
    
    \item \textbf{Space complexity}
        The space complexity, is mainly dominated by two storage needs. First, we need to store all the words being analyzed and their consecutive DLSs. This has a space complexity of $O(n)$. Then we need to store the dominat elements matrix which has a space complexity of $O(g^2)$. Therefore the overall space complexity is $O(n+g^2)$. Therefore, when $g<^2<n$, the space complexity is in $O(n)$.

        The space complexity is mainly dominated by two storage needs. First, we need to store all the words being analyzed and their consecutive DLSs. This has a space complexity of $O(n)$. Then we need to store the dominant elements DLS matrix, which has a space complexity of $O(g^2)$. Therefore, the overall space complexity is $O(n+g^2)$. Consequently, when $g<^2<n$, the space complexity is $O(n)$.
    
\end{itemize}

\subsubsection{Processing logic errors}
\label{sec:LogicErr}
Logic errors in a dataset are the most delicate to identify and rectify. Unlike typographical errors or certain outliers, these errors do not readily reveal themselves, even through observation of the dataset. Logic errors arise when the data does not adhere to the expected logical relationships between variables, making them challenging to detect. To address this issue, we propose an algorithm that, despite its simplicity, effectively identifies these elusive errors. Our method involves the use of data mining techniques to uncover relationships between various variables in the dataset. By identifying and establishing the strongest relationships as true, we can identify observations that fail to conform to these logical connections. Such non-conforming observations are then flagged as inconsistent with the logical structure of the dataset. This approach is used in the work of \citep{CITE_7} and \citep{CITE_8}, where the authors used similar techniques not to specifically detect logical errors but to identify outliers\footnote{In these references, outliers do not designate statistical outliers like in ths work. Instead, they indicate abnormal behavior of a whole record.}. This method ensures that the integrity of the dataset is maintained by validating that the data adheres to the underlying logical relationships expected within the dataset. As we had already mentioned in Section \ref{sec:Abs}, unjustified missing text values will also be flagged in this process.

To detect relationships between variables, we decided to use the association rule mining algorithm, Apriori \citep{CITE_40}. The algorithm is often used in market basket analysis and purchase recommendation systems. Like the Isolation Forest used for outliers, this algorithm seems less known, so we describe its functioning here:

\paragraph{Apriori algorithm}\mbox{}\\
Let us start by recalling the main constituents of the Apriori algorithm:

\begin{itemize}
    \item \textbf{Itemset}: An itemset is a collection of one or more items. A $k$-itemset is an itemset containing $k$ items. 
    
    \item \textbf{Support}: The support is the frequency of a particular itemset. If an itemset $I$ appears in $s$ out of $n$ records, then the support of $I$ is calculated as $s/n$. The minimum support threshold is a user-defined threshold that an itemset's support must meet or exceed to be considered frequent. Itemsets with support below this threshold are discarded from further consideration.
    
    
    
    
    \item \textbf{Association Rule}: An association rule is an implication of the form $I_1 \rightarrow I_2$, where $I_1$ and $I_2$ are itemsets. The rule suggests that records containing $I_1$ are likely to also contain $I_2$.
    
    \item \textbf{Confidence}: Confidence is a measure of the reliability of an association rule. For a rule $I_1 \rightarrow I_2$, the confidence is given by
    $\text{confidence}(I_1 \rightarrow I_2) = \text{support}(I_1 \cup I_2) / \text{support}(I_1)$. The minimum confidence threshold is a user-defined threshold that an association rule's confidence must meet or exceed to be considered strong.
\end{itemize}

The steps of the Apriori algorithm with constituents are as follows:

\begin{enumerate}
    \item \textbf{Generate candidate itemsets}: The algorithm starts with all individual items in the database, generating candidate 1-itemsets. These are evaluated to determine their support.
    
    \item \textbf{Filter candidates}: For each candidate itemset, the algorithm calculates its support. Only those itemsets whose support meets or exceeds the minimum support threshold are considered frequent itemsets. Let us designate by $F_k$, the set of $k$-itemsets kept after the filter.
    
    \item \textbf{Generate frequent itemsets}: Using the frequent $k$-itemsets $F_k$, the algorithm generates candidate $(k+1)$-itemsets by joining $F_k$ with itself. It then filters the candidates that do not meet the support threshold to form the frequent $(k+1)$-itemsets $F_{k+1}$.
    
    \item \textbf{Repeat}: This process is repeated iteratively, generating candidate and frequent itemsets of increasing size until no new frequent itemsets are found.
    
    \item \textbf{Generate association rules}: From the collection of all frequent itemsets, the algorithm generates possible association rules. For each frequent itemset, it generates all possible rules and calculates their confidence. Only the rules that meet the minimum confidence threshold are considered strong association rules.
\end{enumerate}

Naturally, the algorithm in its initial version takes a long time to run as it has a high time complexity. For instance, for each $k$, the algorithm generates $\binom{n}{k}$ candidates. For that reason, we decided to make a change in the functioning of the algorithm. In the original version of the algorithm, the user can adjust the minimum support, as well as the confidence level of a rule. The higher these two values are, the fewer the number of rules that will be searched. We add another parameter that can be adjusted, the maximum length of $k$-itemsets. This gives us the option of limiting the number of elements that can actually define a rule. This affects the number of repetitions conducted in step 4 as well as the sizes of the data being generated and analyzed in steps 1 to 3 of the description above. This change saves a considerable amount of time. Beyond the computational pragmatism of this change, this modification is also supported by the fact that above a certain size of $k$-itemsets, the rule becomes unreadable, and we lose the explainability that is at the core of our concerns in this article.

In our research, we employed the Apriori algorithm with the following parametrization: the minimum support threshold is set at 0.33\% and the confidence level of 99\%. The combination of a low minimum support threshold of 0.33\% and a high confidence interval of 99\% implies that the Apriori algorithm will focus on identifying rare but highly reliable associations within the dataset. This configuration is particularly useful in scenarios where discovering infrequent yet strongly indicative patterns is crucial, such as errors in a data set. By setting a low support threshold, the algorithm ensures that even the quiet uncommon itemsets are considered, while the high confidence requirement ensures that only the most reliable and significant associations are retained. This approach allows for a comprehensive analysis, albeit at the cost of increased computational resources due to the larger number of candidate itemsets. The maximum length of $k$-itemsets is limited at 3, reducing the impact of the low minimum support threshold. This process yields a set of rules, which are subsequently filtered to retain only those with confidence levels below 100\%. In this way, we identify those that have high confidence levels but which in rare cases fail: they are our potential logic errors. For each of these identified rules, all the composing observations (which are a subset of a given record) are marked as invalid due to logic errors. Finally, corrections can be made based on the expected rules that were violated despite the high confidence in those rules.

\paragraph{Complexity analysis}\mbox{}\\
To our knowledge, the literature does not contain research that specifically addresses the issue of the complexity of the Apriori algorithm. Here, we attempt a brief complexity analysis.
\begin{itemize}
    \item \textbf{Time Complexity}\\
    The time complexity of the Apriori algorithm is influenced by the number of records in the database ($n$), the average transaction length ($a$), and the number of frequent $k$-itemsets generated, $C_k$.
    \begin{itemize}
        \item For each level $k$, \textbf{the algorithm generates candidate itemsets} of length $k+1$ from frequent itemsets of length $k$. In the worst case, each pair of frequent $k$-itemsets is considered (has a support above the threshold). The join operation is combinatorial. The complexity of generating candidate itemsets for each level can therefore be approximated as $O(C_k^2)$ in the worst case.
        \item For each candidate $k$-itemset, the algorithm needs to \textbf{scan the entire database to count its support}. Scanning the database $n$ times, takes $O(n a C_k)$ for each level $k$.
    \end{itemize}
    Therefore, the overall time complexity can be expressed as:
    $O\left(\sum_{k=1}^{a} \left(C_k^2 + n a C_k\right)\right)$. Note that in our case, $a$ is bounded to the maximal length of $k$-itemsets that we set at 3.

    \item \textbf{Space Complexity}\\
    The space complexity of the Apriori algorithm is mainly influenced by the number of candidate itemsets stored at each level. At each level $k$, the algorithm stores candidate $k$-itemsets and their support counts. The average space required for this is $O(aC_k)$ for each level. Therefore, the overall space complexity can hence be approximated as: $O\left(\sum_{k=1}^{a} aC_k\right)$.
\end{itemize}

\section{Results}
We recall that for the quality enhancement phase, we have at our disposal (again, without using them) the exact positions of the invalid entries (see Table \ref{tab:errs}). This allows us to compute the exact accuracies we have achieved. However, for the pre-quality enhancement phase, specifically for the primary key, this is not the case.

\label{sec:Results}
\subsection{Pre-Quality enhancement Results}
\subsubsection{Identification of primary keys}
\label{sec:Resprimekeys}
We applied the steps described in Subsection \ref{sec:primkeys}. 

\begin{itemize}
    \item Step 1 of the procedure consisted in identifying, columns with low rates of missing values. This narrowed down the selection to only the twelve out of fifty-three fields that are listed in table \ref{tab:pk1}.

    \item Then, in step 2, identifying fields with specific expressions in their names (\textit{ID}, \textit{CODE}, or \textit{KEY}) would leave us with three fields. Therefore, we need to proceed to step 2.b, which involves finding combinations that allow to identify unique values.

    \item In step 2.b, it is quickly determined that for the subset $K = (SalesID, ModelID)$, only about 0.1\% of value rows are duplicated. Consequently, this subset is identified as the primary key and will be utilized for the subsequent steps of the framework.
\end{itemize}

\begin{table}[H] 
    \centering
    \begin{tabular}{lcc}
        \hline
        \textbf{Field} & \textbf{Variable Type} & \textbf{Proportion of missing values (\%)} \\
        \hline
        \textcolor{red}{auctioneerID}    & \textcolor{red}{INT64}   & \textcolor{red}{0.000} \\ 
        Enclosure       & STR     & 0.025 \\
        fiBaseModel     & STR     & 0.000 \\
        fiModelDesc     & STR     & 0.000 \\
        fiProductClassDesc & STR & 0.000 \\
        \rowcolor{lightblue} 
        \textcolor{red}{ModelID}         & \textcolor{red}{INT64}   & \textcolor{red}{0.000} \\ 
        ProductGroupDesc & STR   & 0.000 \\
        \rowcolor{lightblue} 
        \textcolor{red}{SalesID}         & \textcolor{red}{FLOAT64} & \textcolor{red}{0.047} \\ 
        saledate        & STR     & 0.000 \\
        state           & STR     & 0.000 \\
        YearMade        & INT64   & 0.000 \\
        datasource      & INT64   & 0.000 \\
        \hline
    \end{tabular}
    \caption{Candidates for the primary key. The listed fields have less than 5\% of missing values, the fields in red are potential candidates due to their names and are the first tested in the uniqueness analysis, the fields shaded in blue are the ones finally chosen as the primary key.}
    \label{tab:pk1}
\end{table}
The whole process is executed under 1 second. This efficient execution can be attributed to the fact that we initiated step 2.b with attributes identified during the pattern recognition step. Also note that in a dataset, primary keys are not necessarily a unique subset.

\subsubsection{Processes mapped to specific fields}
\label{sec:MapRes}
After applying the filters introduced in Subsection \ref{sec:Map}, each of the algorithms of the framework will be applied to the following fields.
\begin{itemize}
    \item \textbf{Redundancies} \mbox{}\\
    The redundancy analysis will be conducted on the attributes that comprise the primary key, namely \textit{SalesID} and \textit{ModelID}.
    
    \item \textbf{Absences} \mbox{}\\
    All fields will be screened for missing values.
        
    \item \textbf{Statistical outliers} \mbox{}\\
    The analysis of outliers will only be applied to three fields: \textit{SalePrice}, \textit{YearMade}, and \textit{MachineHoursCurrentMeter}.

    \item \textbf{Typographical errors} \mbox{}\\
    After applying the filter, only three out of fifty-three fields will be analyzed for typographical errors: \textit{state}, \textit{Enclosure}, and \textit{ProductGroupDesc}.

    \item \textbf{Logic errors} \mbox{}\\
    Finally, even after the filtering process, not as many fields are excluded from the analysis compared to the other steps of the framework. Fourteen fields remain: \textit{Enclosure, Ripper, fiBaseModel, Drive_System, state, Transmission, ProductGroupDesc, fiProductClassDesc, Pad_Type, fiSecondaryDesc, saledate, ProductSize, Hydraulics}, and \textit{fiModelDesc}. 
    
\end{itemize}

\subsection{Quality enhancement results}
Overall, the framework we introduce produces reasonably good results. The difficulty of our work lies in finding the right balance between performance and explainability. Any invalid data identified is fully explainable and is natively corrected. The accuracy for statistical outliers and logic errors, could be improved with reasonable ease if we were less exigent regarding the explainability of the results and the algorithms.

In terms of computer performance\footnote{The configuration of the computer used is rather moderate: Intel® Core™ i5-7200U with 2.5 GHz base frequency.}, with the exception of the logic errors detection algorithm, all the algorithms run in a matter of seconds. Handling duplicates takes 1 second, missing values are dealt within less than 1 second, outliers are identified and imputed in 3 seconds, and typographical errors are handled in 33 seconds. The logic error identification took 45 minutes.

\paragraph{Redundancies and Absences}\mbox{}\\
100\% of duplicates and 100\% of missing values were successfully identified. This task does not represent a challenge per se. However, it is interesting to note that the simple rule of thumb used to designate unjustified missing numerical values (95\% threshold) is efficient.

\paragraph{Statistical outliers and Logic Errors}\mbox{}\\
Only 51\% of statistical outliers were identified. As we mentioned in Section \ref{sec:Intro}, detecting statistical outliers while maintaining maximal explainability presents several challenges. The solution we introduced in Subsection \ref{sec:outliers} allows for fully explainable identification. For any flagged variable, it is possible to explain exactly why the variable was considered an outlier. The variable is either too extreme in a non-extreme distribution, or it is isolated from other extreme values in a distribution that has valid extreme values. The issue with this approach is that it does not consider the surrounding variables. For instance, when analyzing statistical outliers in a dataset with fields representing prices and products, it is very challenging to do so without considering the specific products. There are many algorithms that could effectively consider all the surrounding information (typically an Isolation Forest in high dimensions), but they would not be able to flag the specific field with an issue (therefore failing point 1.(c) of our definition of explainability and interpretability) or would somehow not be able to guarantee explainability and interpretability. A solution could be to handle outliers as logic errors. This solution was considered but was discarded due to the heavy computational cost it would imply. Indeed, it would mechanically increase the number of fields to consider for logic errors, and it would require transforming numerical values into categorical ones (for instance, prices from 1 to 100, then 100 to 500, and so on instead of using numerical values).

Only 35\% of logic errors are identified. This result is attributable to the other constraint set in this work, which is not using domain knowledge. Let us recall the logic errors in the dataset introduced in Table \ref{tab:errs}. We had \textit{Wrong category}, \textit{Incoherent machine / Drive system / Product group description}, and \textit{YearMade} > \textit{saledate}. Apart from the last case, all these types of errors could be directly identified using domain knowledge. Specifically, in our case, by using the specific and unique mapping that describes the relationship from one machine (\textit{MachineID}) to its class (\textit{ProductClassDesc}) with all specifications in-between (\textit{Forks}, \textit{Transmissions}, \dots). A solution to keep the framework domain knowledge-free while being more accurate would be to add a third step in the pre-quality enhancement phase that would try to establish dependency relationships between each field of the dataset and then focus an intensive Apriori analysis between these variables without limiting the maximal length of $k$-itemsets. To determine the dependency relationships, solutions that could be explored include, but are not limited to, random forest feature importance \citep{randomforest} and ANOVA analysis \citep{anova}. Additionally, it would be highly beneficial to transform numerical variables into meaningful categorical variables to include them in the logic error detections. As stated above, this would also be advantageous for the detection of statistical outliers. Deepening the research in the mentioned directions would allow for more accurate detection and correction of logic errors while ensuring explainability and interpretability. However, this would come at a hefty computational price. It would be advisable to conduct these evolutions along with parallelized implementation of the algorithms.

\paragraph{Typographical errors}\mbox{}\\
100\% of typographical errors are identified. This is a very good result as we have successfully identified all typographical errors, including entry errors as well as case errors. We have an algorithm that combines clustering and Damerau-Levenshtein distance in an interesting yet not complicated way, achieving remarkable accuracy in detecting typographical errors. The algorithm is made feasible by the simple trick of sorting the words to be corrected alphabetically, which considerably reduces processing (by achieving a very good time complexity, close to $O(n\log n)$, as shown in Subsection \ref{sec:Typos}). For example, for the variable \emph{state}, we start with 102 unique observations among the 100,000 initial words. We end up with 63 groups resulting from the DLS jumps, which corresponds to a reduction of around 60\% in the matrix size of DLSs. Another very good result obtained is that no false positives are produced by the algorithm.

\section{Conclusion}
We have introduced a framework that measures the quality of a dataset without any use of domain knowledge, while maintaining the explainability and interpretability of our results. Our framework is indeed able to identify and correct errors in a dataset without using any information beyond the provided dataset, while being able to fully justify why each flagged observation is considered invalid and propose corrections. The framework handles missing values, duplicates, outliers, typographical errors, and logic errors. After thoroughly reviewing the literature and analyzing our framework in detail, we can see that our work distinguishes itself from the literature by adhering to two constraints (guaranteeing explainability and interpretability and being domain knowledge-free), by the simplicity of the algorithms used, and by the way the algorithms combine machine learning, statistics, and some common-sense tricks.

The analysis of the results we have conducted allows us to draw two main conclusions. First, the concept of analyzing the quality of a dataset without any information about the data will often come at the expense of computing performance. For instance, the need to start the framework with a pre-quality enhancement phase to search for information that would have been obtained in a classical situation is an additional computational cost. Another illustration can be found in the logic error detection algorithm, where domain knowledge would expedite the association rule mining algorithm as the related targets would be manually selected, thus allowing for targeted mining of strong relationships. This would also considerably increase the accuracy of the algorithm. Secondly, the aim of being entirely explainable and interpretable will often reduce the accuracy of the algorithms used. Indeed, this aim systematically involves stepping away from powerful tools, typically some deep learning (DL) and machine learning (ML) algorithms. A good illustration is our choice of using the isolation forest algorithm only in one dimension.
 
Even with these strict constraints, we have also shown that it is entirely possible to obtain very good results. Indeed, apart from outliers and logic errors, all other types of errors have been fully identified and corrected. Typically, the algorithm to identify and correct typographical errors performs very well with a very good complexity. Another conclusion that can be drawn from this exploratory work is that, in a setup constrained as ours, it is necessary to have well-defined pre-analysis steps that will target the analysis and reduce the processing time.

To extend the work we have conducted, many interesting analyses could be undertaken. First, we could consider, as already mentioned, analyzing in more depth the relationships between the attributes. This would not only help reduce computational costs but also provide a better understanding of the database we are working on, allowing for analysis beyond this framework. Additionally, it could be beneficial to consider confidence levels on the corrections made. For instance, when handling outliers, we could run the algorithm multiple times with different parameters, especially $\alpha_s, \; \alpha_k,$ and $\gamma$. Based on the different results, we could explore the possibility of computing a confidence level on the identifications made.

\newpage 

\bibliographystyle{apalike}
\bibliography{references}

\newpage

\end{document}